\documentclass[nofootinbib,longbibliography,prd,floatfix,superscriptaddress,twocolumn]{revtex4-1}
\usepackage[utf8]{inputenc}
\usepackage[hidelinks]{hyperref}
\usepackage[english]{babel}
\usepackage{amsmath,amssymb,amsbsy,amstext,amsthm,booktabs,simplewick,exscale,relsize,graphicx,amsfonts,upgreek,slashed,color,multirow,dcolumn,bm,enumerate,url,todonotes,siunitx,gensymb,titlesec}
\usepackage{geometry}
\usepackage{appendix}
\graphicspath{{figures/}}

\newcommand{\eV}{{\, {\rm eV}}}

\definecolor{mypurple}{RGB}{164,64,214}
\definecolor{mygreen}{RGB}{0,150,50}

\newcommand\thankssymb[1]{\textsuperscript{$\star$}}

\titlespacing*{\section}
{0pt}{2ex}{2ex}
\titlespacing*{\subsection}
{0pt}{2ex}{2ex}

\addto\captionsenglish{}
\begin{document}

\title{New Constraints on Dark Photon Dark Matter with Superconducting Nanowire Detectors in an Optical Haloscope}
\author{Jeff Chiles}
\thanks{Both authors contributed equally to this work.\\ jeffrey.chiles@nist.gov, charaev@mit.edu.}
\affiliation{National Institute of Standards and Technology, 325 Broadway, Boulder, CO 80305}
\author{Ilya Charaev}
\thanks{Both authors contributed equally to this work.\\ jeffrey.chiles@nist.gov, charaev@mit.edu.}
\affiliation{Massachusetts Institute of Technology, 50 Vassar Street, Cambridge, MA 02139, USA}
\affiliation{University of Zurich, Zurich 8057, Switzerland}
\author{Robert Lasenby}
\affiliation{Stanford Institute for Theoretical Physics, Stanford University, Stanford, CA 94305, USA}
\author{Masha Baryakhtar}
\affiliation{Department of Physics, University of Washington, Seattle WA 98195, USA}
\author{Junwu Huang}
\affiliation{Perimeter Institute for Theoretical Physics, Waterloo, Ontario, N2L 2Y5, Canada}
\author{Alexana Roshko}
\affiliation{National Institute of Standards and Technology, 325 Broadway, Boulder, CO 80305}
\author{George Burton}
\affiliation{National Institute of Standards and Technology, 325 Broadway, Boulder, CO 80305}
\author{Marco Colangelo}
\affiliation{Massachusetts Institute of Technology, 50 Vassar Street, Cambridge, MA 02139, USA}
\author{Ken Van Tilburg}
\affiliation{New York University CCPP, New York, NY, 10003, United States}
\affiliation{Center for Computational Astrophysics, Flatiron Institute, New York, NY 10010, USA}
\author{Asimina Arvanitaki}
\affiliation{Perimeter Institute for Theoretical Physics, Waterloo, Ontario, N2L 2Y5, Canada}
\author{Sae Woo Nam}
\affiliation{National Institute of Standards and Technology, 325 Broadway, Boulder, CO 80305}
\author{Karl K. Berggren}
\affiliation{Massachusetts Institute of Technology, 50 Vassar Street, Cambridge, MA 02139, USA}

\begin{abstract} 
Uncovering the nature of dark matter is one of the most important goals of particle physics. Light bosonic particles, such as the dark photon, are well-motivated candidates: they are generally long-lived, weakly-interacting, and naturally produced in the early universe. In this work, we report on LAMPOST (Light $A'$ Multilayer Periodic Optical SNSPD Target), a proof-of-concept experiment searching for dark photon dark matter in the eV mass range, via coherent absorption in a multi-layer dielectric haloscope. Using a superconducting nanowire single-photon detector (SNSPD), we achieve efficient photon detection with a dark count rate (DCR) of $\sim 6\times10^{-6}$ counts/s. We find no evidence for dark photon dark matter in the mass range of $\sim 0.7$-$0.8$ eV with kinetic mixing $\epsilon \gtrsim 10^{-12}$, improving existing limits in $\epsilon$ by up to a factor of two. With future improvements to SNSPDs, our architecture could probe significant new parameter space for dark photon and axion dark matter in the meV to 10 eV mass range.

\end{abstract}
\maketitle

Dark matter (DM), a form of non-relativistic matter
that amounts to $\sim 25\%$ of the energy budget of
the universe \cite{Rubin:1970zza,Aghanim:2018eyx}, is by now the conservative explanation
for a wealth of astrophysical and cosmological data
that cannot be accommodated
within the Standard Model (SM) of particle physics. However, all of our evidence
for DM is via its gravitational interactions on large scales, which is compatible with a very wide range of particle physics models.
 
Light, weakly-coupled new bosons are a well-motivated class of DM candidates~\cite{Dine:1982ah,Abbott:1982af,PRESKILL1983127,Arias:2012az,PhysRevD.93.103520}. Light scalar, pseudoscalar, and vector particles arise in many SM extensions and are generally weakly coupled, long-lived, and difficult to detect~\cite{Svrcek:2006yi,Arvanitaki:2009fg,Holdom:1985ag,Cicoli:2011yh,Dimopoulos:1996kp,Damour:1994zq}. 
These bosonic DM candidates are also automatically produced in the early universe assuming a period of cosmic inflation ~\cite{PRESKILL1983127,PhysRevD.93.103520}. For vector DM, the abundance today depends on the inflationary Hubble scale~\cite{PhysRevD.93.103520}, and can yield the measured DM abundance for DM masses $\gtrsim 5\times 10^{-5} \eV$ given current constraints on the inflationary scale~\cite{Akrami:2018odb}. The detection of a  vector DM particle at the eV scale would point to a Hubble scale of $5\times 10^{12}$~GeV, otherwise unreachable in any laboratory experiment or late-universe astrophysical observation.

The simplest and least-constrained vector DM model is the dark photon, characterized by the `kinetic mixing' interaction with the photon~\cite{Holdom:1985ag},
\begin{align}
	\mathcal{L} \supset &-\frac{1}{4} F_{\mu\nu} F^{\mu\nu}
	- \frac{1}{4} F'_{\mu\nu} F'^{\mu\nu}  
	- \frac{1}{2} \epsilon F_{\mu\nu} F'^{\mu\nu} \nonumber \\
	&+ \frac{1}{2}m_{A'}^2 A'^2
\end{align}
where $F_{\mu\nu}$ and $F'_{\mu\nu}$ are the field strengths of the photon and the dark photon, respectively, and $A'$ is the dark photon field.
The dark photon mass $m_{A'}$ and kinetic mixing parameter $\epsilon \ll 1$ define the DM parameter space.

Similarly to a photon, the leading interaction between dark photon DM and a detector is the
\emph{absorption} of the DM particle \cite{PhysRevD.98.035006, PhysRevX.8.041001}. 
The entire rest mass energy $m_{A'}c^2$ can be captured, in contrast to scattering, which deposits at most the kinetic energy $m_{A'} v^2/2$ in direct detection experiments \cite{Goodman:1984dc} (where $v \sim 10^{-3}c$ is the galactic DM velocity and $c$ is the speed of light). This motivates new experimental schemes for detection of light bosonic DM. In this work, we focus on the efficient conversion of dark photon DM to near-IR photons.

\begin{figure}[t]
    \centering
    \includegraphics[width=.9\columnwidth]{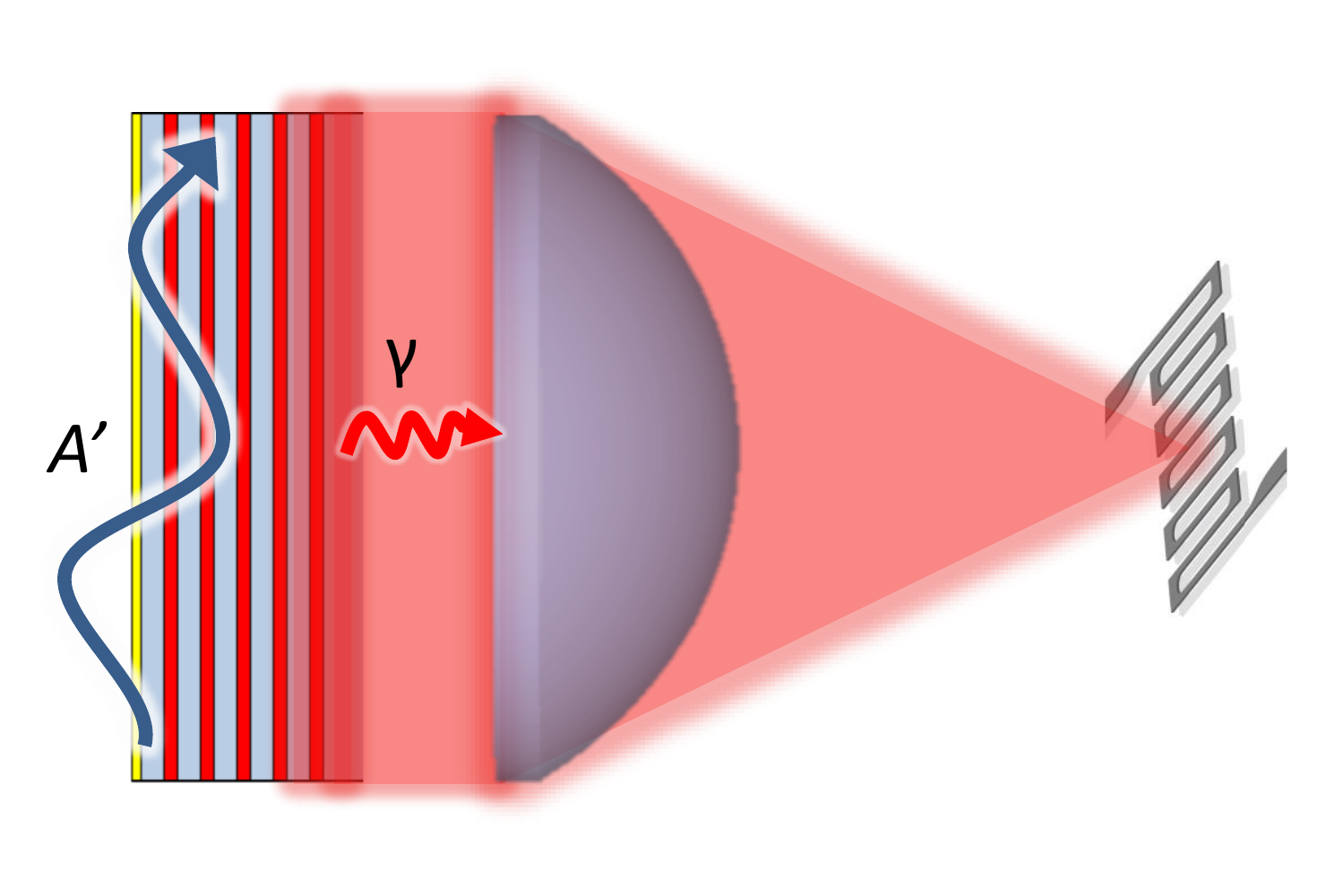}
    \caption{Sketch of the LAMPOST concept.
	The dark photon dark matter field $A'$ converts
	to photons in a layered dielectric target.
	These photons are focused by a lens onto a small, low-noise SNSPD detector. The beam emitted from the stack is approximately uniform except for a small region in the middle where a mirror is absent, not shown here.}
	\label{figcartoon}
\end{figure}

To convert a non-relativistic dark photon into
a relativistic photon of the same frequency,
the target must compensate for the mismatch in
momentum. 
This can be achieved using a stack of dielectric layers
with different indices of refraction, whose
thicknesses are on the scale of the photon's wavelength~\cite{PhysRevD.98.035006,TheMADMAXWorkingGroup:2016hpc,orpheus,etiger1, etiger2};
we use a stack of half-wavelength layers~\cite{Baryakhtar:2018doz}. 
In such a structure, dark photon DM at the corresponding
frequency can convert coherently to photons: the photon acquires its energy from the dark photon DM and its momentum from the lattice vector of the photonic crystal, thus alleviating the momentum mismatch between the non-relativistic DM and the relativistic photon (see Fig.~\ref{figcartoon} for a sketch
of the setup). In particular, in a half-wave stack \cite{Baryakhtar:2018doz}, due to constructive interference between converted photons from different layers, the conversion rate increases as the square of number of layer periods $N$.
If the half-wave frequency is matched to the DM mass, 
then the converted power per unit area is given by
\begin{equation}
	\frac{P}{A} \simeq \frac{8}{3} \epsilon^2 \rho_{\rm DM} N^2 \left(\frac{1}{n_1^2}
	-\frac{1}{n_2^2}\right)^2 ,
\end{equation}
where $\rho_{\rm DM} \simeq 0.4 {\rm \, GeV \, cm^{-3}}$ is the local DM density,
$n_{1,2}$ are the refractive indices of the alternating layers,
and $A$ is the area of the stack
(Fig.~\ref{pfig1} shows the converted power as a function
of DM mass for our stack configuration).

Due to the small DM velocity, the converted photons are emitted within $\sim 10^{-3}$ rad of the normal vector to the layers. This allows them to be focused down to an area $\sim 10^{-6}$ smaller than that of the layers, permitting the use of small, highly sensitive detectors \cite{PhysRevD.98.035006,PhysRevX.8.041001,TheMADMAXWorkingGroup:2016hpc}. Superconducting nanowire single-photon detectors
(SNSPDs) have demonstrated, in separate experiments, ultralow
dark count rates ($10^{-6} {\rm \,
Hz}$) necessary to detect rare signal events, active areas large enough to collect the
focused light ($\gtrsim 0.1 {\rm \, mm^2}$),
near-unity detection efficiency, and sensitivity
to photons from 0.1 eV to 10 eV \cite{PhysRevLett.123.151802, Reddy20,
verma2020singlephoton, Wollman17}. These properties make SNSPDs well-suited to the unique requirements of this project.

In this work, we present the first results from the LAMPOST (Light $A'$ 
Multilayer Periodic Optical SNSPD Target) experiment with 180 hours of data collection. 
Our simple and inexpensive prototype constrains new dark photon DM parameter space at masses $\sim 0.7$-$0.8$ eV (corresponding to photon wavelengths $\sim$ 1550-1770 nm) with less than a week of run time. The choice of the $\sim 0.7$-$0.8$ eV mass range allows us to leverage off-the-shelf equipment and robust fabrication processes to simplify this proof-of-concept experiment.

\textit{Experimental setup.}\textemdash The dielectric stack, or target, generates the signal photons of interest. As discussed in \cite{PhysRevD.98.035006}, a useful configuration
is a `half-wave' stack, in which the stack's layers have alternating refractive indices  $n_1,n_2,n_1,n_2,\dots$ and thicknesses $d_1,d_2,d_1,d_2,\dots$, with $n_1 d_1 = n_2 d_2$. The thicknesses and indices are chosen for light at the signal wavelength of interest to acquire $\pi$ phase upon transmitting through each layer. In such a material, dark photon DM with frequency $\omega \simeq \pi/n_i d_i$ can convert coherently
to photons. We utilise alternating layers of amorphous silicon and silica, deposited on top of a $\sim 0.525 {\rm \, mm}$ thick silica substrate which is polished on both sides. 

\begin{figure}[t]
    \centering
    \includegraphics[width=1\columnwidth]{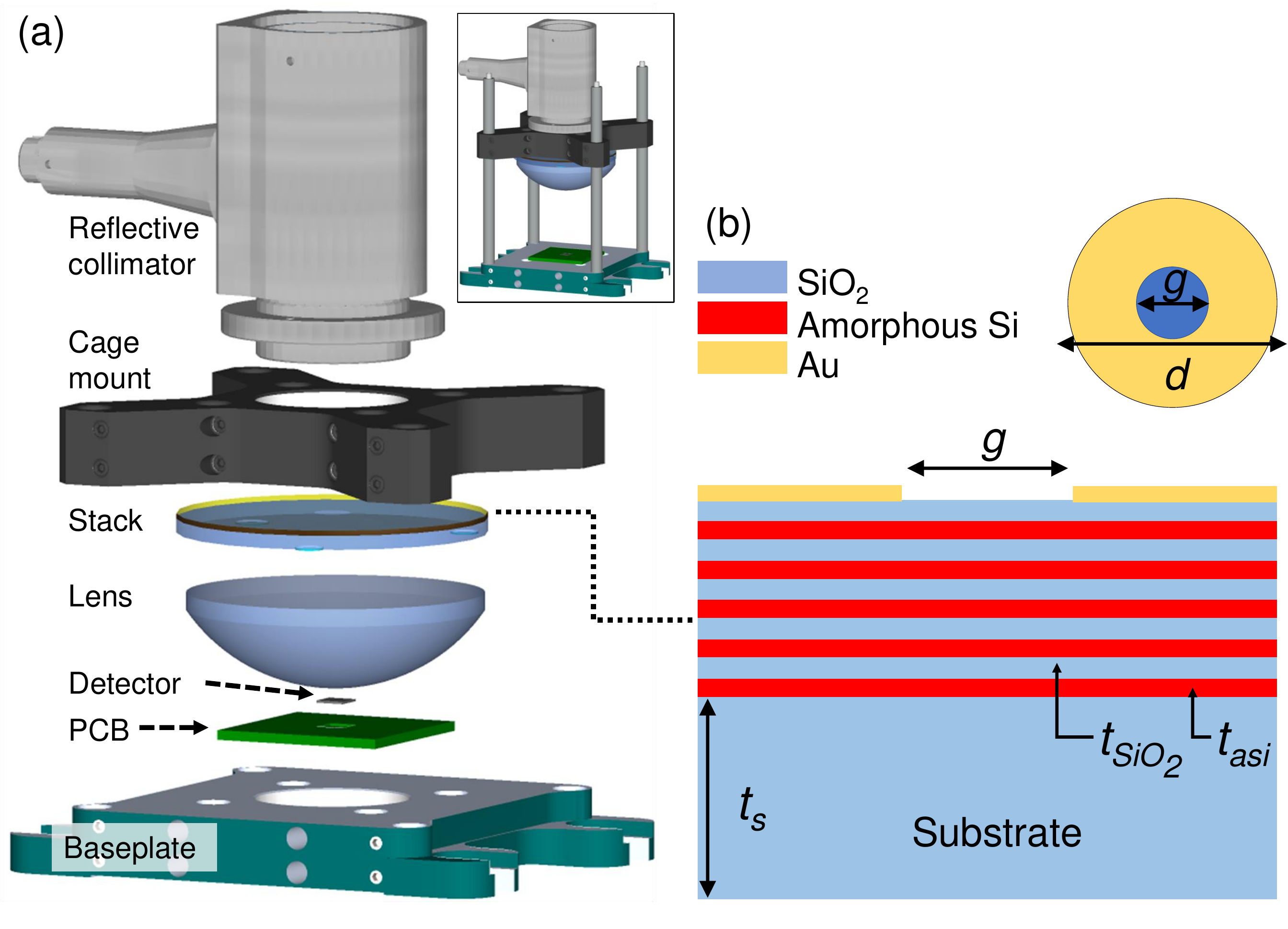}
	\caption{The LAMPOST prototype haloscope apparatus.  (a) Exploded view with element details. Inset: assembled view. b) Schematic cross-sectional and top-views of the dielectric stack target responsible for DM-signal photon conversion, with designed values of different dimensions, \textit{g}: Aperture diameter, 10 mm; \textit{d}: Wafer diameter, 50 mm; \textit{t\textsubscript{s}}: Substrate thickness, 525 $\mu$m; \textit{t\textsubscript{asi}}: Amorphous silicon layer thickness, $\sim$292 nm; \textit{t\textsubscript{s}}:~SiO\textsubscript{2} layer thickness, $\sim$548 nm. See Supplementary Materials for details of the film characterization.}
    \label{fig:apparatus}
\end{figure}

\begin{figure}[t]
\centering
\includegraphics[width=\columnwidth]{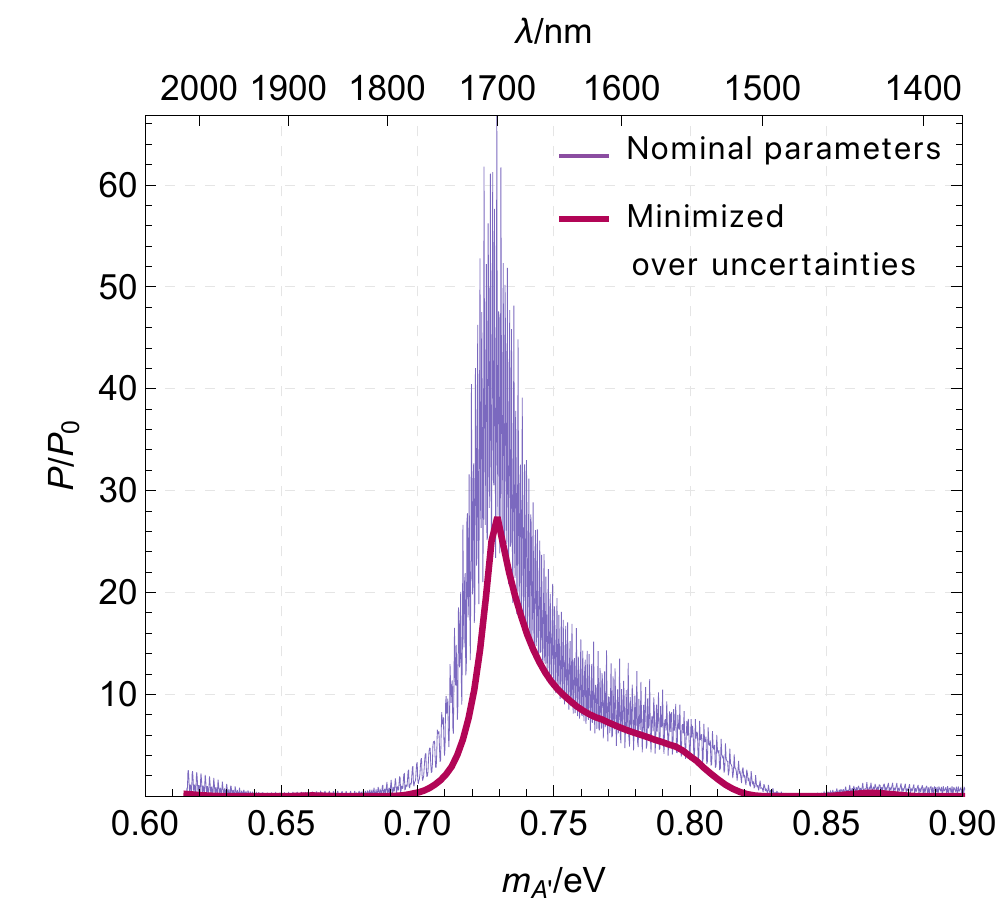}
\vspace{-.6cm}
    \caption{Calculated time-averaged power $P$ absorbed from dark photon DM with mass $m_{A'}$ by the
	layered target, normalised to the power $P_0$ absorbed by
	a uniform mirror. The thin purple curve shows
	the power absorbed by a target with parameters given by
	 their respective measured central values,
	while the magenta curve shows the minimum power
	obtained by varying the parameters within 
	measurement uncertainties (see Supplementary Materials). The substrate
	thickness is assumed to physically vary by $\gtrsim 10 {\rm \, \mu m}$
	over the target area; this accounts for the magenta
	curve sometimes falling above the purple curve of constant
	substrate thickness.}
    \label{pfig1}
\end{figure}

The dielectric stack is integrated and aligned with several optomechanical elements and a 50 mm focal length plano-convex lens to focus the signal onto the primary
SNSPD. The structure is illustrated in the right-hand panel
of Fig.~\ref{fig:apparatus}. A reference SNSPD, nominally identical to the first, is placed on the same PCB as the primary detector, but offset by 2 cm so as to be completely out of the optical path of the signal. This reference detector can serve to provide an estimate of the event rate for several potential sources of background counts, including cosmic ray muons, Cherenkov photons generated in the lens \cite{du2020sources}, or high energy particles excited by radioactive decay events. The entire apparatus is contained inside a light-tight box.

To electrically and optically characterize the fabricated SNSPDs, we designed an experimental setup using a sorption-pump type He-3 cryostat. The haloscope containing the SNSPDs (the assembly in Fig. \ref{fig:apparatus}) was placed on a 300 mK cold stage, with ample spacing from the innermost radiation shield. The signal was amplified
at the 4 K stage by a cryogenic low-noise amplifier with a total gain
of 56 dB and then was sent to a pulse counter. A single mode optical fiber
delivered light from 1550 nm and 1700 nm CW lasers into the cryogenic apparatus
though a vacuum feedthrough. 

An important consideration is whether the apparatus is mechanically stable enough to preserve the intended alignment during the cooldown. A mock haloscope was constructed to independently test this (which used the same SNSPD and PCB).  We can detect a misalignment by comparing the experimental and theoretical DE of the SNSPD; a large discrepancy would suggest substantial misalignment. First, we placed a conventional, large-area optical power meter directly above the SNSPD at room temperature and recorded the optical power at a fixed laser power output at 1550 nm.  Next, we removed the power meter, cooled the system to 300 mK, and recorded the photon count rate on the SNSPD for the same laser output power (but with a fixed and known optical attenuation added to the signal path to avoid saturating the SNSPD). A DE of 28.3$\pm$ 0.5\% was observed for the case of light polarized along the length of the wire (parallel) and 12.1$\pm$ 0.5\% for the perpendicular polarization. A paddle-type polarization controller was used to shift the state as needed.  We simulated the theoretical DE of our detector to be 33.6\% for the parallel polarization case and 10.6\% for the perpendicular case (described further in the next section). We note that the experimental parallel DE of 28.3$\pm$ 0.5\% could be 1.117 times higher, or 31.6\%, if the SNSPD were operated at a higher bias current (this is necessary to facilitate direct comparison to the simulations, where the internal detection efficiency is assumed to be unity).  Comparing the experimental DE of 31.6\% (after compensating for incomplete saturation of the internal detection efficiency) to the simulated DE of 33.6\% for the parallel polarization case, the magnitude differs only by a factor of 0.94.  The discrepancy may be explained by small temporal variations in laser source power ($\sim$2\%),  incomplete polarization state purity ($\sim$1-2\%), variable scattering loss at fiber connectors at different temperatures, and an acceptable amount of misalignment in the beam.  The larger-than-expected perpendicular polarization DE and smaller-than-expected parallel DE would be consistent with a slightly impure polarization state during the measurement. We note that the targeting beam changed position by 100 $\mu$m in both the lateral directions upon warming up after this mock haloscope test, which was consistent with the behavior of the main haloscope during the actual experiment.

Separately, the main haloscope was assembled and tested briefly prior to the data collection. A brief optical measurement at cryogenic temperatures was conducted (without control over polarization), giving a reasonable DE of 19.3\% which is between the nominal parallel and perpendicular DE values. Next, the system was warmed to room temperature, and the optical fiber was disconnected to prevent blackbody radiation-induced counts impinging on the detector.  After cooling down again, we began recording counts on the main detector over several cycles of the cryostat.  At several points, the haloscope was removed from the system, and the optical alignment was inspected to ensure no significant drift had occurred.  A translational drift of about 100 $\mu$m was observed, consistent with the detailed alignment test conducted separately. We collected count data for both the main and reference SNSPDs over a total time of 180 hours at the $300 {\rm \, mK}$ base temperature, while operating both SNSPDs at a bias current of 4.2 $\mu$A.

After assembling the dielectric stack and the SNSPD as in Fig.~\ref{fig:apparatus}, we combine simulated and experimentally measured factors to obtain a well-bounded number for the system detection efficiency (SDE), which captures all known sources of loss from the point of signal photon generation in the stack to the generation of photon count events in the SNSPD. The SDE can be expressed as
\begin{align}
    SDE = OCE \cdot T \cdot DE.
\end{align}
The optical collection efficiency (\textit{OCE}) is derived from ray-tracing simulations as described in the Supplement. The simulations show that 1.27\% of signal photons generated in the stack impinge on the SNSPD in the worst-case misalignment. The OCE constitutes the largest source of loss in our system, and is limited by several factors, including a $\pm 100 {\rm \, \mu m}$ in-plane alignment uncertainty (experimentally observed), spherical aberration, and total internal reflection losses in the lens.  Additionally, we found that wafer curvature resulting from intrinsic stress in the dielectric stack's thin films modified the focal length from the expected value, further misaligning the signal (see Supplement). The transmission coefficient $T=88\%$ captures a small optical loss incurred by defects in the dielectric stack which scatter the signal. Finally, the detection efficiency \textit{DE} of the SNSPD is the probability of generating a detection event for one photon incident on the detector's footprint; the value is estimated to be 17.5\% based on a calibrated measurement of the DE at 1550 nm, which is averaged for both polarization states, followed by an adjustment for 1700 nm photons (the detector is roughly 10\% less sensitive at 1700 nm).  Overall, we achieve an SDE of 0.20\% in our system. The calculated converted power $P$ per unit target area $A$,
as a function of dark photon mass, is shown in Fig.~\ref{pfig1}.
This power is normalised to the time-averaged power converted by a simple mirrored surface, $P_0/A = \frac{2}{3} \epsilon^2 \rho_{\rm DM}$, where $\rho_{\rm DM}$ is the local DM energy density.

\begin{figure}[t]
\centering
\includegraphics[width=\columnwidth]{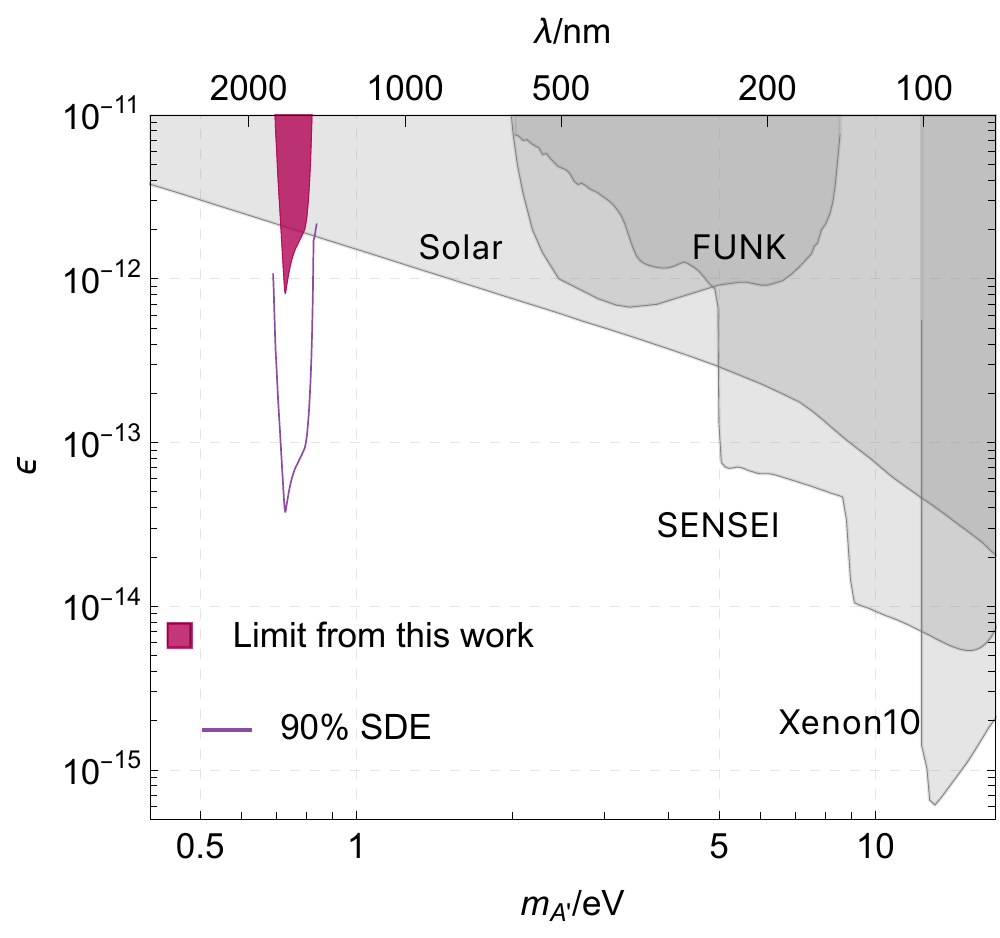}
    \caption{LAMPOST constraints on dark photon DM with mass $m_{A'}$ and kinetic mixing $\epsilon$. The magenta shaded region
	shows the $90\%$ limit set by our experiment.  The thin purple curve corresponds
	to the reach of an equivalent experiment with an improved SDE of $90\%$.
	Existing limits on dark photon DM
	from the FUNK~\cite{PhysRevD.102.042001},
	SENSEI~\cite{PhysRevLett.125.171802} and Xenon10~\cite{PhysRevLett.107.051301} experiments and from the non-detection of Solar dark photons
	by Xenon1T~\cite{PhysRevD.102.115022} are shown in gray.}
	\vspace{-.6cm}
    \label{kfig1}
\end{figure}

\textit{Results.}\textemdash Over a 180 hour exposure, the primary SNSPD registered 4 counts,
while the reference SNSPD registered 5 counts. As discussed in the Supplementary Material, the dark count
rates for the two SNSPDs are likely to be similar. Accordingly, we can estimate
the dark count rate for the primary SNSPD using the reference SNSPD's counts,
and set a limit on the kinetic mixing $\epsilon$, as described in the Supplementary 
Material. 
Figure~\ref{kfig1} shows this 90\% confidence limit,
derived by minimizing over measurement uncertainties
as described below,
compared to existing bounds~\cite{PhysRevD.102.115022}.
We assume a local DM density of $0.4 {\rm \, GeV \, cm^{-3}}$,
with a standard truncated Maxwell-Boltzmann velocity distribution
\cite{10.1103/PhysRevD.82.023530}, and assume that the dark photon polarization direction varies randomly over timescales longer than the DM coherence time (see Supplementary Materials for details).

There are several measurement uncertainties on the
properties of the target, such as the layer thicknesses. To set conservative limits,
we calculate the minimum signal power that is
compatible with the possible range of target properties 
(as discussed in the Supplement). This procedure has a non-negligible effect,
since the large substrate thickness
$t_s \simeq 525 \pm 10 {\rm \, \mu m}$
introduces an oscillatory dependence of the signal
power on the DM frequency. The time-averaged DM absorption rate per unit area,
as a function of dark photon mass, is shown in Fig.~\ref{pfig1}.
The thin purple curve shows the signal power 
for a target with coplanar, uniform layers,
with the measured central value thicknesses and properties, illustrating the rapid oscillation caused by the large substrate thickness.
Small variations in target properties
can these oscillations by more than a period, introducing uncertainty in the signal power at a given DM mass.
Analogously, physical variation in e.g.\ the substrate
thickness over the disk (slowly varying over the span of the wafer, known as total thickness variation) results in a disk-averaged
conversion power that is averaged over shifted
curves. Taking both of these effects into account
gives the magenta curve in Fig.~\ref{pfig1}.

The calculations for Fig.~\ref{pfig1} were
carried out in 1D (using transfer matrix
methods~\cite{1612.07057}), treating
layers locally as infinite and uniform.
As discussed above and in the Supplement,
there will be deviations from this approximation.
However, these deviations will only be important if,
at scales $\lesssim$ the DM coherence length
(which is $\simeq 0.5 {\rm \, mm}$ at the relevant DM masses),
they have a greater effect than the uncertainties incorporated
into the 1D calculation.
As discussed in the Supplement,
measurements of the stack show that, with
the exception of small defects (which were incorporated
into our loss calculations above), the layer
properties are uniform enough that
the uncertainty in converted power is dominated by the
1D effects. Another effect is that, due to wafer
curvature, the angle between the layers
and the lens surface will vary over the stack,
being effectively coplanar in some places but not in others.
Taking the minimum over 1D configurations which include
the lens and its anti-reflection coating (treating them as coplanar
with the stack), and those which do not, does not visibly alter the magneta
curve in Fig.~\ref{pfig1} --- the uncertain substrate thickness
is already a large effect. 
The misalignment between the stack and the lens
will also alter the focussing of the stack-generated
signal, which is taken into account by our optical collection
efficiency calculations.
While the anti-reflection coating,
which is itself a series of dielectric layers (whose
properties we do not know precisely),
could hypothetically convert DM into a photon signal
which may interfere with that from the stack, 
the variation in stack-lens separation over the area,
and the fact that the lens and stack are separated
by more than the DM coherence length, means that
the converted powers will add incoherently
(this is discussed in more detail in the Supplement).

Fig.~\ref{pfig1} illustrates that,
even using our conservative estimates, our layered
target enhances the conversion rate by up to a factor $\sim 30$ times that of a
mirror target of the same area~\cite{Horns:2012jf,Suzuki:2015sza,PhysRevD.102.042001}.
Compared to dark photon absorption in the SNSPD itself,
as considered in~\cite{PhysRevLett.123.151802},
the much larger area and multiple layers of the dielectric target produce a signal rate from stack-converted photons that is at least $\sim 10^3$ times greater (at its optimum frequencies).  Fig.~\ref{kfig1} shows that our prototype detector constrains previously unexplored DM
parameter space in the mass range $0.7-0.8 \eV$.

As discussed in the Supplementary Materials, since
we do not know the precise cause of our observed counts, it
is possible that the dark count rates for the reference
and primary SNSPDs are somewhat different.
To set a conservative limit on the kinetic mixing $\epsilon$,
we can find the value of $\epsilon$ that
would lead to $\le 4$ signal counts only $10\%$ of the time,
giving a $90\%$ confidence limit on the coupling,
independent of any dark count rate estimate.
This results in a $\epsilon$ limit a factor $\sim 1.3$ times
larger
than the nominal limit shown in Figure~\ref{kfig1}.
In future experiments, further measurements
(e.g.\ a control experiment where the stack is removed)
would enable more precise measurement of the
primary SNSPD's dark count rate. 

\textit{Discussion.}\textemdash There are several clear directions toward extending the experimental reach beyond the prototype. The first is improving the OCE from its current value of 1.27\%, which suffers from a combination of optical aberrations and reflective losses. In the future, these could be mitigated with a longer-focal length lens, reaching 93\% OCE (see Supplementary Fig. 4), but at the expense of mechanical stability. With custom-designed adapters to house the lens, stack and collimator, and finer toleranced parts used in the assembly, concerns over alignment could be assuaged. The other aspect would be improving the DE of the SNSPD. By adding an appropriately designed dielectric coating around the SNSPD, the DE could be raised to 98\%\cite{Reddy20}. Finally, by lowering the defect count of the target's dielectric stack, transmission losses could be made negligible. Combining these would enable an SDE above 90\%, which would increase the reach of an otherwise equivalent experiment about an order of magnitude in coupling, as illustrated in Fig.~\ref{kfig1}.

Further improvement can be achieved from background count characterization and mitigation, to determine whether they originate from cosmic ray muons, Cherenkov photons generated in the lens~\cite{du2020sources}, or simple statistical fluctuations of the bias current in the detector.  For dark counts generated by cosmic rays or radioactivity, it may be possible to veto such events using additional detectors. 

Extending the reach to heavier DM masses could be accomplished with wider-bandgap thin film materials such as ZnSe or TiO\textsubscript{2}.  To search lighter DM masses, more research should be conducted on fabricating low-energy threshold SNSPDs, though promising results have been obtained at photon energies as low as $\sim 0.1 \eV$ ($\lambda \sim 10 {\rm \, \mu m}$)
\cite{verma2020singlephoton,marsili2012efficient}.

By placing the dielectric layers in a magnetic field,
axion \cite{axion1,axion2,axion3} DM with a coupling to photons \cite{10.1146/annurev-nucl-102014-022120} could also be absorbed,
allowing a haloscope to probe axion masses well above the 
traditional microwave range~\cite{PhysRevD.98.035006}. 
If good SNSPD performance in a large
magnetic field is achieved
(as has been demonstrated in some
cases~\cite{POLAKOVIC2020163543,lawrie2021multifunctional}), almost the same experimental setup could be used. 

The LAMPOST prototype places the first constraints on dark matter using optical haloscopes, exceeding current constraints in the $0.7-0.8$~eV mass range by up to a factor of two in dark photon coupling. At the same time, the prototype demonstrates technologies and techniques that will enable searches over even larger volumes of parameter space. Optimizing the optical collection and detection efficiency of the setup can improve the coupling limits by more than an order of magnitude.
Larger volumes of layered dielectric targets, longer integration times, parallel operation of complementary frequency haloscopes, and background characterization and vetoes are all concrete avenues toward a rapid exploration of large regions of dark photon dark matter parameter space. Integration with a large background magnetic field and lower-threshold SNSPDs will enable the search for axion dark matter in the meV mass range.

RL's research is supported in part by the National Science Foundation under Grant No.~PHYS-2014215, and the Gordon and Betty Moore Foundation Grant GBMF7946.
Some of the computing for this project was performed on the Sherlock cluster at
Stanford. We would like to thank Stanford University and the Stanford Research Computing Center for providing computational resources and support that contributed to these research results.
We thank Dr.\ J.\ Hilfiker of J.A.\ Woollam Corp. for providing VASE measurements. We thank Daniel Egana-Ugrinovic, Rouven Essig, Alexander Millar, Kent Irwin, and Saptarshi Chaudhuri for useful comments and discussions. Research at Perimeter Institute is supported in part by the Government of Canada through the Department of Innovation, Science and Economic Development Canada and by the Province of Ontario through the Ministry of Colleges and Universities. The MIT co-authors acknowledge support for the later stages of the work from the Fermi Research Alliance, LLC (FRA) and the US Department of Energy (DOE) under contract No. DE-AC02-07CH11359. The initial stages of the work were supported by the DOE under the
QuantiSED program, Award No. DE-SC0019129. The MIT co-authors also thank Brenden Butters for the technical work.

\nocite{Moharam:81}
\nocite{pedrotti2017introduction}
\nocite{Lyon1977ThermalExpSi}
\nocite{Hahn1972ThermalExpSio2}
\nocite{vlassak}
\nocite{Stoney1909}
\nocite{wikipedia_2021}
\nocite{Komma2012ThermoOptic}
\nocite{2105.04565}
\nocite{Supp}

\newpage

\appendix
\clearpage

\section*{Supplementary Materials}

\section{Construction of the apparatus}
\label{construction}
The haloscope ``core'' comprises an aluminum 50~mm optical cage mount,
the dielectric stack wafer, and a plano-convex focusing lens epoxied
together, with a reflective collimator threaded onto the backside. The
core is mounted to Invar posts to minimize thermal contraction, and the
posts are mounted to an aluminum baseplate which houses the SNSPD's PCB.
The entire haloscope is mounted and housed in a light-tight aluminum
enclosure to minimize stray light entering the cryostat through various
feedthroughs at higher temperature stages.

The detector is permanently mounted at the three dimensional location of best focus considering the position and orientation of the dielectric stack and the lens.  This location is determined by propagating a 635 nm laser from the reflective collimator all the way to the detector. The gold reflector is omitted from an aperture of width $g = 10$ mm in the dielectric stack wafer, allowing the laser to pass through.
This allows it to simulate the propagation vector of the DM signal photons.
The SNSPD is translated within the plane of the baseplate, and the
haloscope core is moved along the focal axis until the alignment beam is
visibly focused and centered on the detector area, as viewed through an
off-axis microscope and camera setup. Next, the haloscope core is moved
closer to the detector by a pre-calculated amount that compensates
differences in source properties (wavelength, beam size, beam shape)
between the alignment beam and DM signal photons. Finally, the SNSPD is
fixed in place with UV-cured epoxy.

\section{Optical simulations and calculations}
\label{optical_simulations}
In this experiment, two major aspects require optical simulations to evaluate the ideal and experimental efficiencies of the system.  The first is the optical system (spanning the point of signal photon generation in the stack, to the detector plane), and the second aspect is the absorption on light in the SNSPD. We will describe our methodology and results in turn for each aspect.

We chose to analyze the optical system with non-sequential ray tracing, which efficiently captures the effects of many aberrations and misalignment effects present in the physical system.  The desired quantity is the optical collection efficiency (OCE) which captures all losses in the optical system. First, it is useful to understand the relationship between the alignment beam (to point the dielectric stack at the detector) and the DM signal that would hypothetically be produced inside the stack.  The aperture in the stack gives an annulus source shape for the DM signal and a circular beam profile for the alignment beam.  Because of spherical aberrations and chromatic focal shift, these two beams do not share the same focal distance.  In the laboratory setting, the alignment beam can be observed to pass through a focus on the plane of the detector by moving the haloscope core (lens, stack, collimator).  If we replicate the system parameters (lens shape, source properties such as profile, wavelength, etc) in a ray tracing model, we can also obtain an equivalent position of best focus, shown in Supplementary Fig. \ref{fig:alignmentmethod}(a) as \textit{z\textsubscript{align}}.  It need not be in exactly the same position as the experimentally observed focal distance.  Next, in the ray tracing model, we change the source properties to mimic the expected DM signal, which is an annulus with a beam diameter of \textit{d} = 50 mm and a circular stop the same diameter as the aperture \textit{g} = 10 mm, emitting at 1550 nm (Supplementary Fig. \ref{fig:alignmentmethod}(b)).  The new position of best focus for the DM signal is found at a displacement of \textit{dz\textsubscript{signal}}, or -1 mm (negative values being closer to the detector plane) in our case.  The effect of thermal contraction in the invar cage rods supporting the haloscope core was also included in this number. Knowing \textit{dz\textsubscript{signal}}, we can use it during the experiment by first locating the visible focus for the 635 nm alignment beam, then moving the haloscope core closer to the detector by 1 mm.

Separately, we considered several physical effects on the dielectric stack itself that may influence the optical signal in our experiment.  During cooling, the a-Si thin films will contract differently from the SiO\textsubscript{2} substrate, resulting in stress and wafer bowing.  We used temperature-dependent coefficients for the linear thermal expansion coefficients of a-Si and SiO\textsubscript{2} \cite{Lyon1977ThermalExpSi,Hahn1972ThermalExpSio2} and calculated the stress resulting from the contraction from the deposition temperature (40$\degree$C) to 10K (the difference between 10K and the working temperature of 300 mK being insignificant), via \cite{vlassak}

\begin{equation*}
\begin{aligned}
 \sigma = E' \Delta\bar{\alpha} \Delta T,
\end{aligned}
\end{equation*}
where $E'$ is the biaxial modulus of silicon (180 GPa), $\Delta\bar{\alpha}$ is the average difference in thermal expansion coefficient between silicon and fused silica over the entire temperature range of consideration, and $\Delta T$ is the change in temperature, which is 303 K.  This results in a tensile stress of 57 MPa.  The resultant induced wafer curvature was calculated via the Stoney formula \cite{Stoney1909} as $-47.4$ m, resulting in an optical focusing power $1/R = -0.021$ m\textsuperscript{-1} for photons generated in the stack.  Adding this to the optical power of the lens, the effective focal length of the stack-lens system increases by about 0.05 mm, which we consider to be negligible given the low sensitivity to focal length and the wide tolerance range already assumed.

We also considered another mechanism by which the optical focus of the signal could be perturbed, which is intrinsic stresses of the thin films in the dielectric stack.  We measured the thin film stresses in reference films of a-Si and SiO\textsubscript{2} to be -160 and -240 MPa, respectively.  Considering these to act independently in the limit of small strain and curvature, we recalculated the resultant wafer curvature (in this case 4.7 m) for the thin films in our stack with the Stoney formula above and added them to the optical power of the lens, similar to the treatment for the thermal film stress.  We find that the effective focal length decreases by 0.5 mm, which introduces a systematic offset in focal distance for data collected during the experiment, since it was not compensated for at the time of alignment and data collection.  In the following section describing ray tracing results, the signal source is adjusted to mimic this initial focusing effect.

\begin{figure}[t]
\includegraphics[width=\columnwidth]{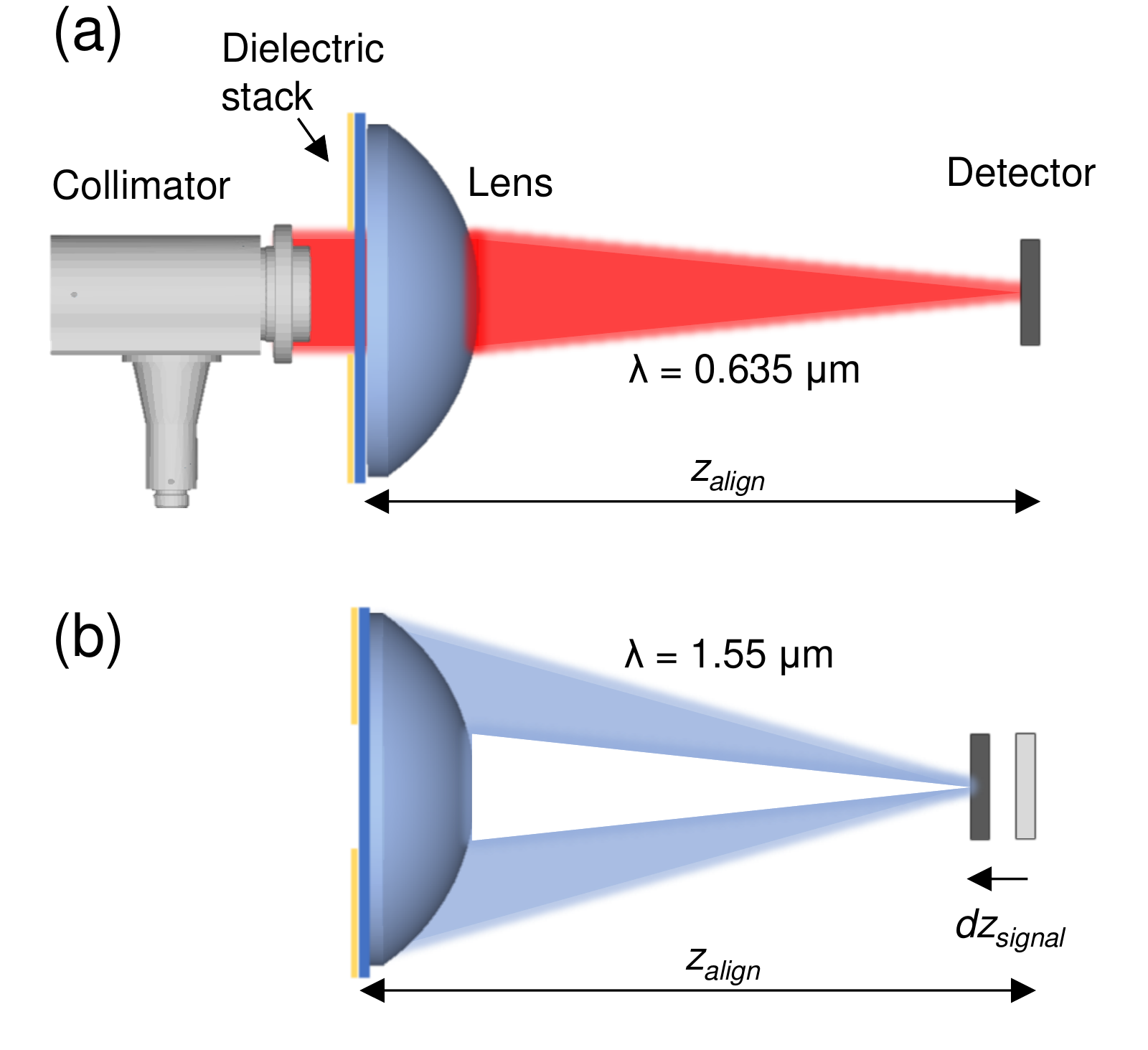}
\caption{Optical system used in ray-tracing and alignment planning.  (a) Propagation of the red alignment beam through the aperture, resulting in a position of best focus \textit{z\textsubscript{align}} for the detector during the initial alignment. (b) Effective behavior of the DM signal photon source as an annulus with a modified position of best focus ($\ z_{align}-dz_{signal}$).}
\label{fig:alignmentmethod}

\end{figure}

\begin{figure}[h]
\includegraphics[width=\columnwidth]{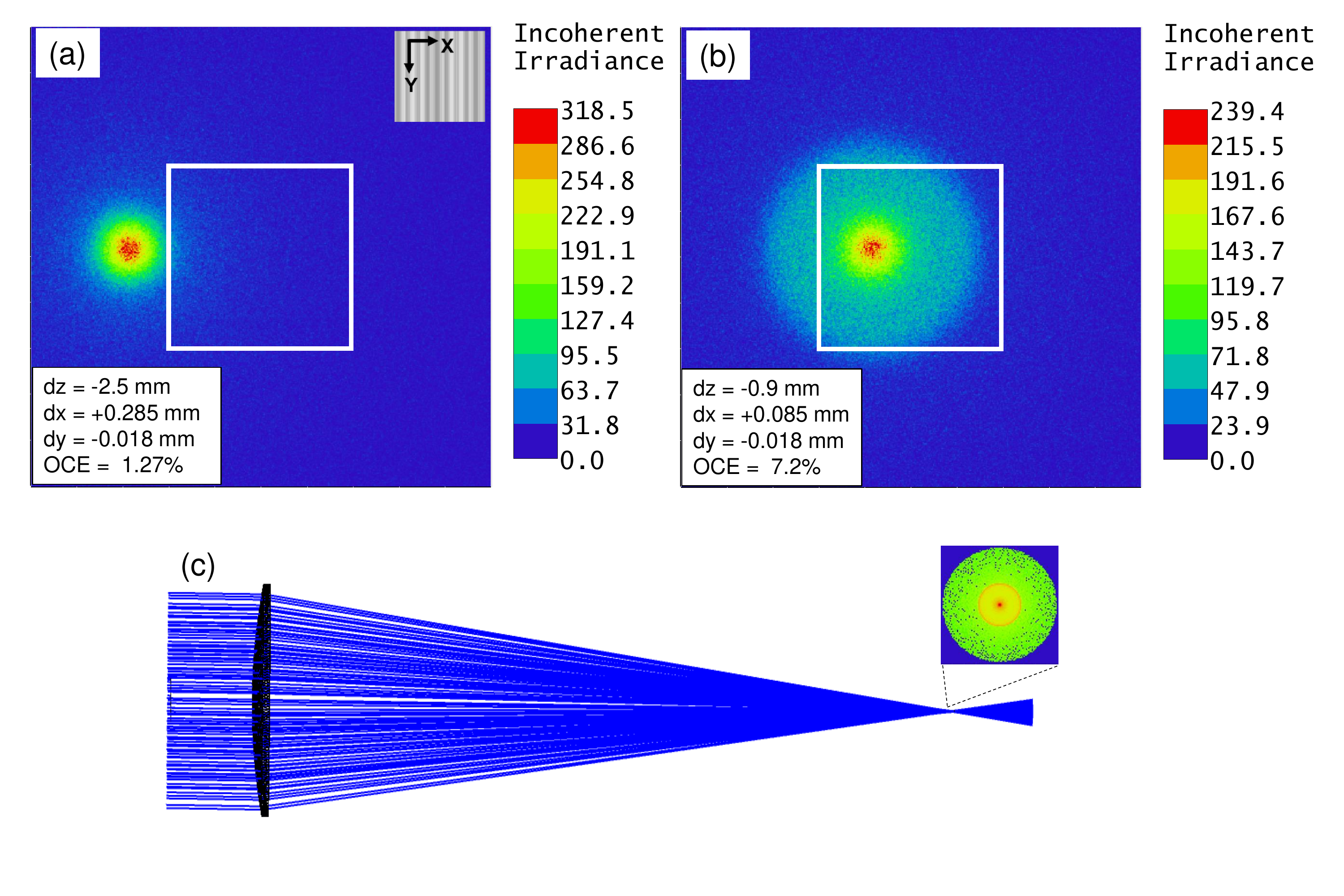}
\caption{Non-sequential ray tracing results for the (a) worst-case misalignment scenario and (b) the best-case scenario, showing incoherent irradiance (linear scale color mapping) at the plane of the detector in each case. The detector's active area is indicated by the white boxes. Inset of (a): coordinate system used for displacements. (c) Ray tracing of the improved optical design using a longer focal length lens which achieves 93\% OCE. Inset: Incoherent irradiance in log scale at the plane of the detector in the case of the improved optical design.}
\label{fig:raytracing}
\end{figure}

Ray tracing is used to calculate the estimated OCE for the system when it is aligned at this position, for the peak signal wavelength of 1700 nm (note this differs from the expected wavelength of 1550 nm due to larger-than-intended layer thicknesses). We identified two sources of misalignment that are sufficiently large to consider in this estimation, which are defocus (deviation from the best focal length in the \textit{z} axis) and unwanted detector translation along the \textit{xy} plane away from the calculated DM signal spot position.  We estimate a defocusing uncertainty of $\pm$~2.5 mm in our experiment.  The translation error constitutes a fixed and known misalignment contribution as well as a systematic uncertainty.  The fixed misalignment itself comes from (i) off-axis tilt in the reflective collimator, causing the alignment beam position to differ from the DM signal position by a small amount, and (ii) an initial error in the detector position when it was glued down, due to shifting during the curing of the epoxy used to fix it in place.  Both these values were summed and used in the ray tracing simulations shown in Supplementary Fig. \ref{fig:raytracing} as a starting location for the DM signal beam's position relative to the detector.  We noted that the alignment beam's position in the plane of the detector was observed to vary by $\pm$ 100~$\mu$m when inspected between cooldown cycles.  As there is no reason for this error to vary during data collection once the apparatus has cooled to $\sim$300 mK, and since we cannot inspect the alignment beam's position when cooled down, we incorporated this as a systematic uncertainty to be tested for worst-case estimation of the OCE.  Ray tracing simulations were performed to compare several plausible misalignment scenarios.  We varied the beam position by $\pm$ 100~$\mu$m in either in-plane axis and by $\pm$ 2.5 mm along the focal length and found the worst case scenario of possible misalignment outcomes, in which the detector is shifted relative to the signal beam by +100~$\mu$m along the \textit{x}-axis, and -2.5 mm on the \textit{z}-axis sending much of the beam off the surface at its most tightly focused point (Supplementary Fig. \ref{fig:raytracing}(a)). The OCE for this scenario is calculated to be 1.27\%.  In Supplementary Fig. \ref{fig:raytracing}(b) we also considered the best-case misalignment outcome, which gives an OCE of up to 7.2\%. We note that displacement in the \textit{y}-axis does not significantly affect any of the results, within the tolerance observed. $4 \times 10^6$ rays were used for non-sequential ray-tracing.  To simulate the effect of the DM signal's angular spread (due to DM velocity dispersion), the rays were scattered with a Gaussian angular distribution (standard deviation of 0.85 mRad angular deflection per ray away from the normal vector). The lens was coated in simulations with the exact AR-coating that was applied to the real lens in the experiment.  As a final note, we neglected the influence of signal light transmitted through the aperture portion of the stack, where the reflector is absent, as it is expected to be weak in comparison to the rest.  

We also note the ray tracing results for a modified haloscope optical path (Supplementary Fig. \ref{fig:raytracing}(c)) which uses a plano-convex lens of a much larger focal distance (200 mm) with the convex side facing the stack.  These changes would minimize the effects of spherical aberration and total internal reflection losses, resulting in a significantly improved OCE of 93\%.

Earlier, we mentioned an anti-reflection (AR) coating applied to the lens in the haloscope.  In our case, the coating is proprietary to the manufacturer of the lens and designed for broadband operation transmission within a certain spectral window (like the near-infrared, in our case), or a so-called broad-band anti reflection (BBAR) coating.  Such a coating may consist of numerous alternating dielectric layers, typically arranged in a numerically optimized, aperiodic set of thicknesses. Since the number of layers and their composition are unknown to us, one may consider whether the BBAR coating could generate its own signal capable of destructively interfering with the dielectric stack's own signal.  In order for this to occur, several conditions must be satisfied: (1) the BBAR coating must be very specifically designed to concentrate a similar amount of power into the exact spectral window in which we are operating, (2) it must be almost perfectly co-planar across most of its surface, and specifically positioned to produce destructive interference, and (3) the dielectric stack and the BBAR coating must be less than a coherence length apart, in order for the signal powers to interfere at all. As discussed earlier in this Supplement, (2) is not satisfied, since wafer bow causes the gap between the stack and the lens to be significantly non-uniform, which would result in averaging of the signals over a large area, rather than perfect destructive or constructive interference.  Additionally, small and slow thickness variation across the stack wafer introduces another effectively random perturbation to this distance.  Moreover, (3) is not satisfied since the dielectric stack is at least one coherence length ($\lambda_c \simeq 0.53 {\rm \, mm} \, (0.7 \eV / m_{A'}$) away from the BBAR coating on the lens, considering the thickness of the stack substrate (525 $\mu$m), plus contributions from the gap between the lens/stack from the epoxy used to attach them, and the additional gap due to wafer bow.  Finally, we conducted ray tracing to consider a final aspect relevant to this issue: due to the wafer bow, a ray originating from a specific point on the BBAR coating and another ray originating from the same relative location on the dielectric stack are consistently displaced from one another along the focal plane of the detector (due to the effective local tilt of the stack wafer adding a defocus term to the BBAR coating's signal).  As a result, the two signals would be significantly misaligned on the detector, again preventing perfectly uniform interference.  Since multiple conditions for uniform destructive interference from the BBAR coating are not satisfied, we conclude that it can be safely neglected from our experiment.  However, in the future, it is worth considering that a deliberately designed and deposited BBAR coating on the lens could be an economical means of obtaining a thick and highly uniform dielectric stack in itself, without the need for a separate stack wafer.

We also conducted optical simulations of the SNSPD's detection efficiency under various conditions for the purpose of validating our alignment strategy and confirming nominal operation of the SNSPD.  We conducted a combination of 2D finite-element modeling and rigorous coupled-wave analysis \cite{Moharam:81} to obtain this value, using the best available measurements of material thicknesses, refractive indices, and lateral dimensions of the as-fabricated SNSPD.  Considering only the light incident from the top surface of the SNSPD, the theoretical detection efficiency is 32.7\% for the parallel polarization (and 10.2\% for the perpendicular polarization).  However, given the large detection area of this device and the size of the near-infrared beam used for alignment, the reflection off the backside of the silicon substrate cannot be neglected in the calculation of the total absorption coefficient in the detector. This factor is most significant when the beam is centered on the detector area but focused on the bottom surface of the substrate.   Conservatively, if we assume the worst case in which the focal spot of the beam is centered on the detection area and focused on the bottom side of the substrate, the light transmitted through the SNSPD will diffusely scatter off the unpolished back silicon surface with a Lambertian intensity profile.  The total absorption coefficient of this backward pass can then be calculated by integration of light scattered to angles encompassing the SNSPD's active area, and multiplication by the angle-dependent absorption coefficient of the SNSPD for light incident from underneath, referred to as the backward pass.  For simplicity, we neglect light scattered into the perpendicular polarization, which has a much smaller absorption coefficient. Light scattered under a Lambertian profile has a radiant intensity proportional to the cosine of the angle between the incident beam and the surface normal \cite{pedrotti2017introduction}. We start with the calculation of radiant flux $F_{tot}$ emitted from a Lambertian surface \cite{wikipedia_2021}:

\begin{equation*}
\begin{aligned}
 F_{tot} = 2\pi I_{max} \int_{b}^{c} \frac{\sin(2\theta)}{2} d\theta ,
\end{aligned}
\end{equation*}
where $\theta$ is the angle of incidence, $b$ and $c$ are the angular span of integration, and $I_{max}$ is is the peak radiant intensity chosen as $1/\pi$ so that $F_{tot}=1$ when integrating $\theta$ from 0 to $\pi/2$, the full angular range available in reflection.  To calculate the angular dependence of the total absorption in the detector during the backward pass, we include the angular dependence of the detector's absorption coefficient in the integral of the previous expression, and integrate over the angles of incidence in which the detector is visible to the point of reflection:

\begin{equation*}
\begin{aligned}
 \alpha_{bkwd} = 2\int_{0}^{\Omega} \frac{\sin(2\theta)}{2} \alpha(\theta)d\theta ,
\end{aligned}
\end{equation*}
where $\alpha_{bkwd}$ is the total absorption coefficient of the backward pass, $\alpha(\theta)$ is the angle-dependent absorption coefficient of the detector from backward incidence (around 21\% for relevant angles, obtained from 2D simulations), and $\Omega$ is the maximum angle of incidence where the detector area is still visible to the reflected beam (0.49 rad). We calculate $\alpha_{bkwd}$ to be 4.9\%. Next, we consider that only some of the original light incident on the top surface remains at the point of reflection off the back surface.  The transmission coefficient through the detector, $T$, is 0.62, and the reflection coefficient $R$ off the back surface is estimated to be uniformly 0.306, which when multiplied together and with $\alpha_{bkwd}$ result in an effective increase in the DE of about 1\% due to the backward pass.  Adding this to the forward-pass DE, the total detection efficiency of the SNSPD is estimated to be 33.6\% in the case of an incident beam perfectly aligned to the detector area, and polarized parallel to the nanowire's long axis. For the case of the perpendicular polarization, the backward pass has a marginal influence, increasing the total DE from 10.2\% to about 10.6\%.  We ignore the effect of light scattered into the rest of the substrate, since the 1 cm\textsuperscript{2} die is large enough that most of the captured light will escape before encountering the relatively small detector area again.

\begin{figure}[t]
    \centering
    \includegraphics[width=0.8\columnwidth]{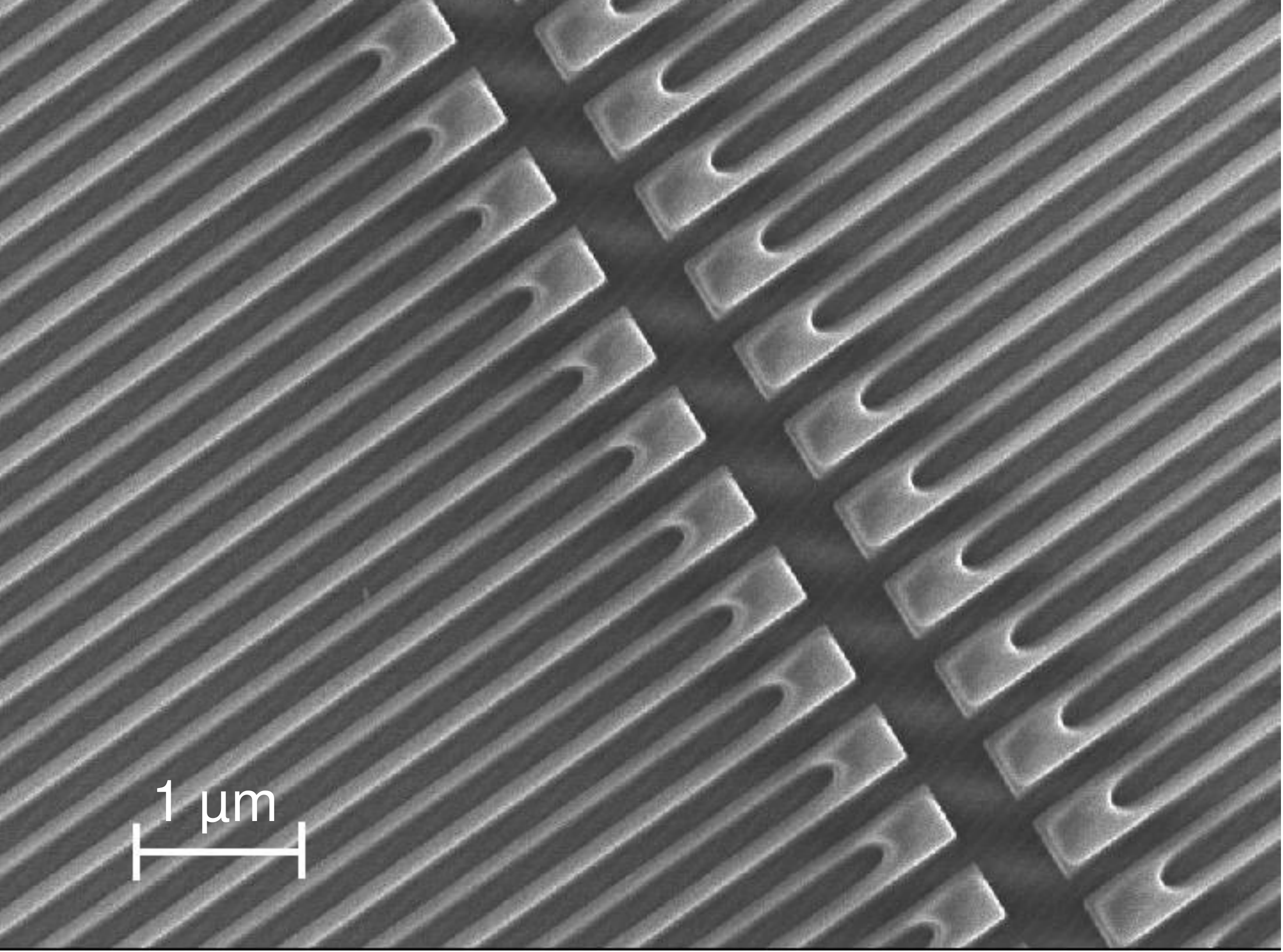}
    \caption{Scanning electron micrograph of nanowires after fabrication.}
    \label{fig:SEM}
\end{figure}

\section{Dielectric stack fabrication and characterization}
\label{stack_fab}
The dielectric stack, which is responsible for resonant conversion of DM particles to signal photons, was fabricated at NIST Boulder.  The starting substrate was a fused silica wafer 50 mm in diameter, which was mounted with wax to a 150 mm-diameter carrier wafer at the position of best uniformity previously recorded for the deposition system to be used.  Prior to film coating, it was plasma-cleaned in oxygen.  The dielectric stack was deposited with alternating films of hydrogenated amorphous silicon (a-Si) and silicon dioxide (SiO\textsubscript{2}) deposited at 40\degree C with inductively-coupled plasma chemical vapor deposition (ICP-CVD).  Based on measured deposition rates and prism-coupled refractive index measurements, the stack was designed to consist of 5 pairs of 292 nm a-Si and 548 nm SiO\textsubscript{2}.  A gold reflector with a 5 nm Ti adhesion layer was deposited on top of the last SiO\textsubscript{2} layer with electron beam evaporation. 
\begin{figure}[h]
\includegraphics[width=\columnwidth]{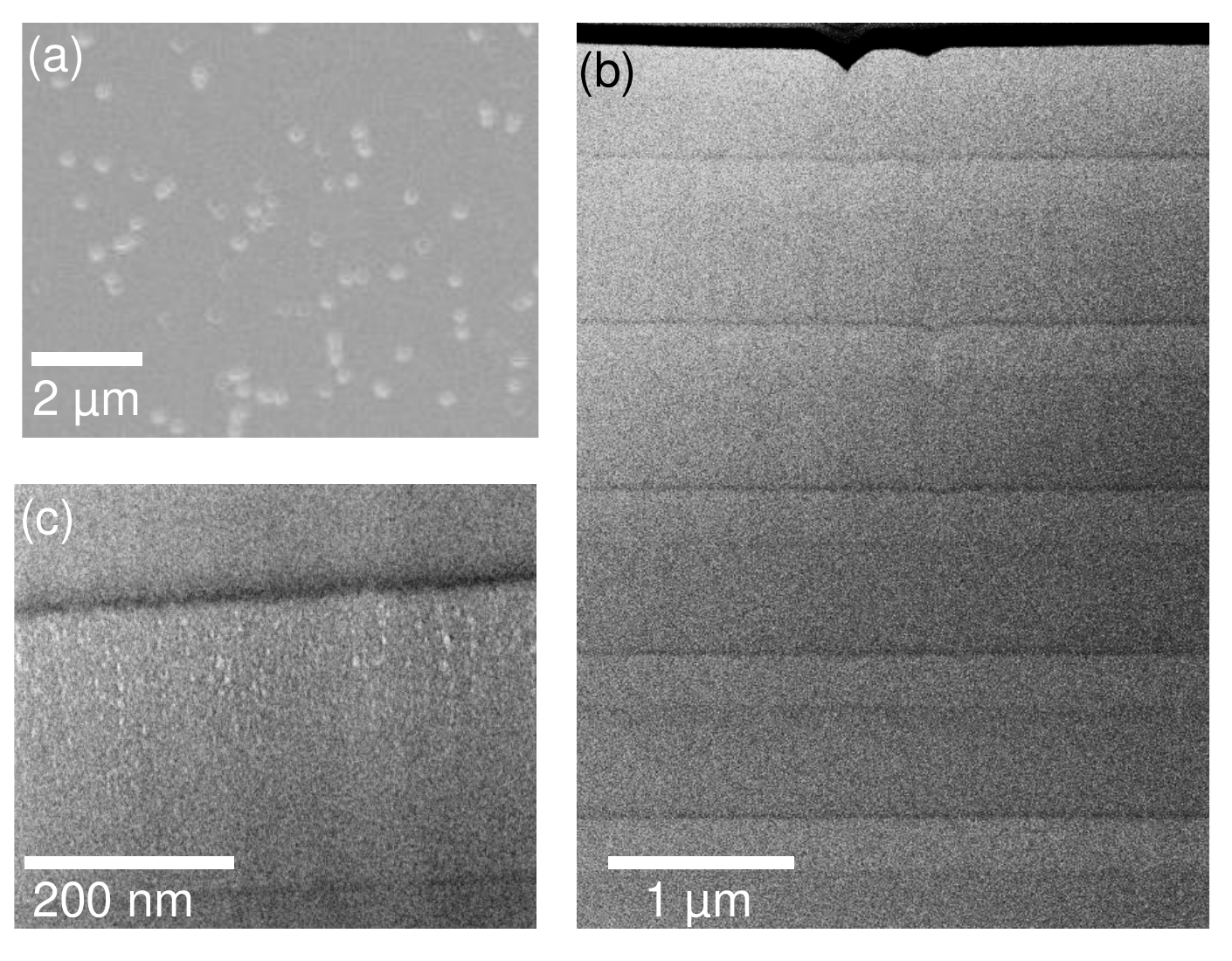}
    \caption{STEM and SEM images of the fabricated stack. (a) Top-view showing numerous small pits occupying a small portion of the surface area. (b) Image of the entire stack, showing one pitted area extending through several layers. (c) High-magnification TEM image of one amorphous silicon layer.}
    \label{fig:stack_stem}
\end{figure}

\begin{figure}[h]
\includegraphics[width=\columnwidth]{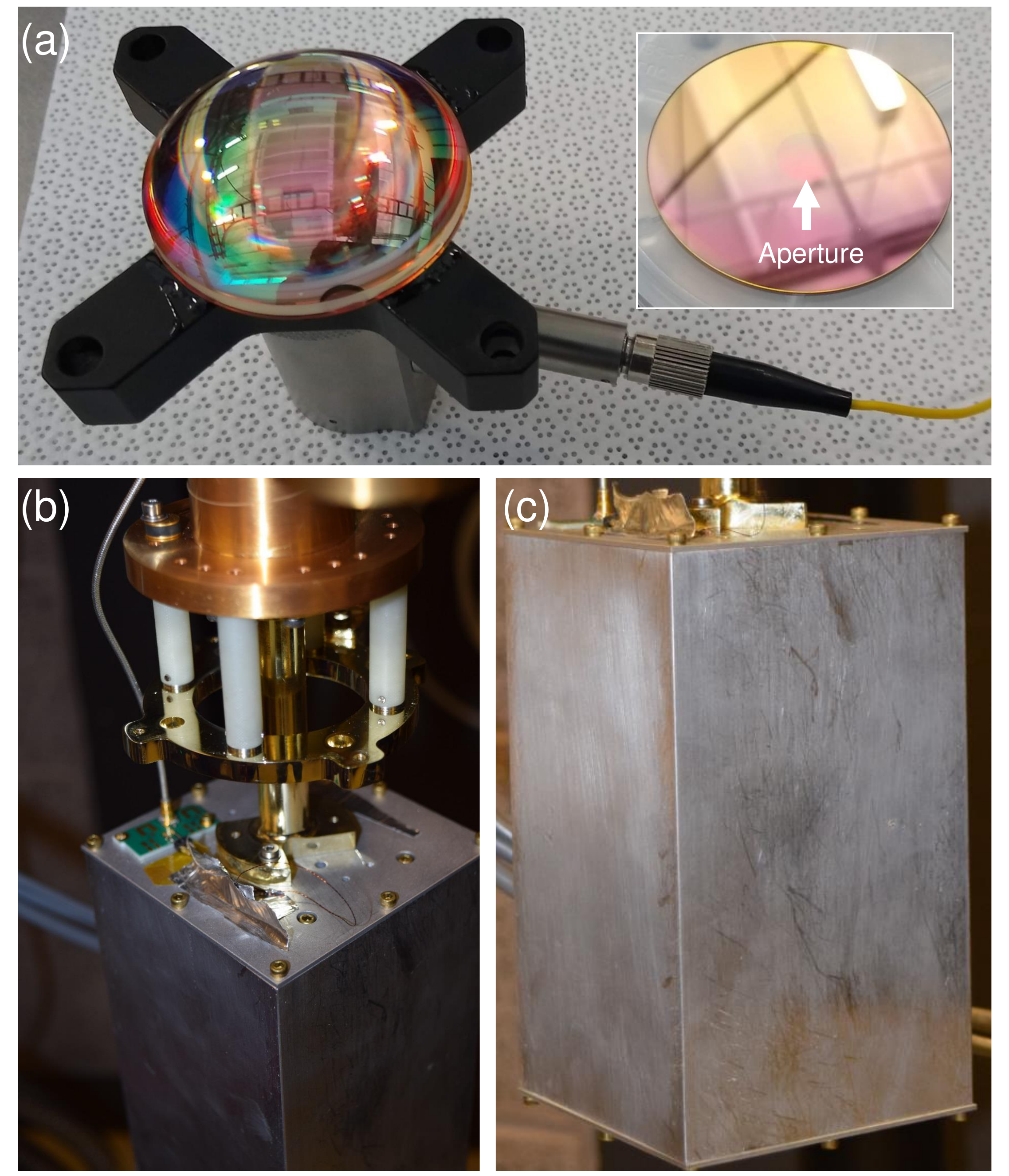}
    \caption{(a) Assembled haloscope core with fiber collimator. Inset: backside of dielectric stack prior to core assembly. (b) A haloscope enclosure mounted on the 300 mK stage of cryostat. (c) View of light-tight enclosure.}
    \label{fig:box}
\end{figure}

Supplementary Fig.~\ref{fig:stack_stem} shows several images acquired through scanning transmission electron microscopy (STEM) and scanning electron microscopy (SEM) of one sample of the fabricated stack. Supplementary Fig. \ref{fig:stack_stem}(a) is a top-view image showing the presence of numerous small (0.1 $\mu$m\textsuperscript{2} area) pits present on the sample.  These do not typically extend down to the bottom layer (as seen in Fig. \ref{fig:stack_stem}(b)), but nevertheless we assume that no signal is generated from the area of the stack intersecting with any blisters, resulting in a stack transmission coefficient of 0.88 (or 12\% loss) which factors into the final SDE calculation.  Small interface effects were also observed between layers, which may be a consequence of sputtering damage or interdiffusion during the early stages of each layer's deposition.  

The precise layer thicknesses were determined with variable-angle spectroscopic ellipsometry over a wavelength range from 200 - 1700 nm.  The refractive index of each type of film (such as a-Si or SiO\textsubscript{2}) was fitted over all measured data but assumed to be constant for all layers.  Layer thicknesses were left uncoupled to allow for interlayer thickness variability. Experimentally, we observed actual thicknesses of 590 nm, 591 nm, 593 nm, 591 nm, and 595 nm for the SiO\textsubscript{2} layers (starting at the gold reflector and moving toward the substrate), and 289 nm, 282 nm, 286 nm, 280 nm, and 288 nm for the a-Si layers in the same order.  Considering two samples from different areas on the wafer, a variability of $< \pm$ 0.2\% was observed, which has a negligible effect on the stack's behavior.  The overall bias toward thicker-than-intended SiO\textsubscript{2} layers may have resulted from different deposition behavior for the fused silica wafer which was wax-mounted to a silicon carrier wafer.  The near-infrared refractive index of the a-Si layer was determined to be 2.64, and the refractive index of the SiO\textsubscript{2} layer was 1.48, both measured at room temperature. Both indices are constant over the mirror's spectral response range. The imaginary components of the refractive indices were previously measured to be $< 10^{-4}$ in the wavelength range of interest to this experiment, making them negligible for the optical performance of the stack. While the optical properties of the stack are well known at room temperature, the a-Si may encounter some degree of thermo-optic shift during cooling to the 300 mK base temperature. We used the temperature dependence of the thermo-optic coefficient for crystalline silicon as a model \cite{Komma2012ThermoOptic} and determined that the a-Si would drop to a refractive index of 2.61 (and the SiO\textsubscript{2} would shift by a negligible amount).  This has been factored into calculations for signal power generated by the stack.

\begin{figure}[t]
    \centering
    \includegraphics[width=0.98\columnwidth]{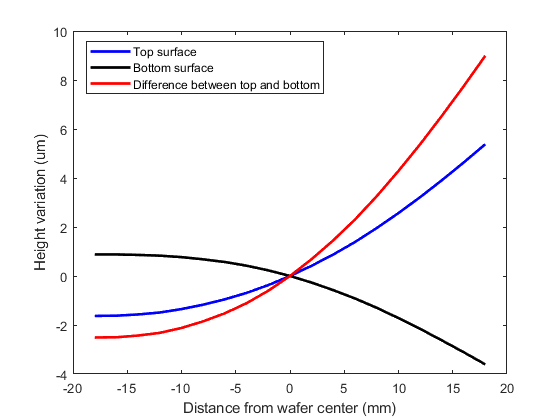}
    \caption{Laser surface topography measurement results of a typical fused silica substrate from the same batch of wafers used for the dielectric stack.} 
    \label{fig:waferheight}
\end{figure}

We also characterized the properties of the stack's substrate, which is a 50 mm diameter fused silica wafer with a nominal thickness of $525 {\rm \, \mu m}$. In the context of this experiment, we are interested to know the surface topography and thickness variation of the wafers. Since DM has a coherence length of $\simeq 0.53 {\rm \, mm} \, (0.7 \eV / m_{A'})$, and so signal photon generation is coherent over this length, it is important that the signal wavefront is not disturbed by a large amount over this span.  Since the original stack wafer was destructively tested for precision measurements of the stack earlier, we analyzed several substrates from the same batch to capture the worst-case topography variations that could be expected.  The wafers were scanned in two ways.  The first way uses an array of ~mm size laser beams reflected off the surface and imaged at a distance to precisely measure wafer curvature and topographical variations over large areas.  For a given wafer, it was scanned in a line across one axis, then was flipped upside down, and the scan repeated on the bottom surface in the same axis.  The height variation in each case was taken relative to the middle of the wafer.  This provides information on the relative variation of the substrate surfaces across its area.  The top surface height variation is plotted in Fig. \ref{fig:waferheight}.  The height is seen to vary by about 7 $\mu$m across the entire surface, which is several times smaller than the magnitude of surface height deformation induced by wafer bow from intrinsic film stress.  At its steepest slope, this is less than 0.2 $\mu$m deviation per coherence length, which is a negligible amount.  It is worth noting that no leveling is done on this height data, so we also plotted the difference between the top and bottom surface height variation as the red line in Fig. \ref{fig:waferheight}, showing little difference in the magnitude of the deviation.

To be certain that there were no high-spatial-frequency variations in surface topography, we also conducted 3D laser confocal scans of the surface in an area roughly as large as the coherence length.  The relative surface height deviations were considerably smaller than the signal wavelength over the entire region, for several random regions measured on a given wafer.  Thus, we conclude that the stack substrate surface height uniformity is sufficiently good for this experiment's needs.

Taking all of these measurements in account, we obtained
a conservative lower bound on the converted power,
displayed in Fig.\ 3 of the main text, by minimizing the signal
power over plausible values for the different target
parameters. Specifically, we scanned the layer
thickness over a $\pm 5 {\rm \, nm}$ range
around their measured values, and the substrate thickness
over a range $\pm 10 {\rm \, \mu m}$ around its nominal
value. The refractive indices for the SiO$_2$ layers
were scanned over a $\pm 0.01$ range, while
those for the Si layers were scanned over $\pm 0.05$.

\section{SNSPD fabrication and testing}
\label{snspd_fab}
The device was fabricated from 7-nm thick WSi film which was sputtered on a 150 nm thick thermal silicon oxide film on a silicon substrate at room temperature with RF co-sputtering. Additionally, a thin 2-nm Si layer was deposited on top of the WSi film in-situ to prevent oxidation of the superconductor. To pattern the nanowires, electron-beam lithography was used with high-resolution positive e-beam resist. The ZEP 520 A resist was spin coated onto the chip at 5000 rpm which ensured a thickness of 335 nm. After exposure, the resist was developed by submerging the chip in O-xylene for 80 s with subsequent rinsing in 2-propanol. The ZEP 520 A pattern was then transferred to the WSi by reactive ion etching in CF\textsubscript{4} at 50 W for 5 minutes.  The ZEP thickness is estimated to be 250 nm after etching and is left on the top surface.

Supplementary Fig.~\ref{fig:SEM} is a scanning electron micrograph (SEM) of the 
tungsten silicide (WSi) SNSPD after fabrication. The device
area was 400 by 400~$\mu$m$^{2}$, and the nanowire was connected to
external circuitry via two contact pads. The width of the nanowires was 140 nm with a pitch of 340 nm. 

\begin{figure}[t]
    \centering
    \includegraphics[width=0.98\columnwidth]{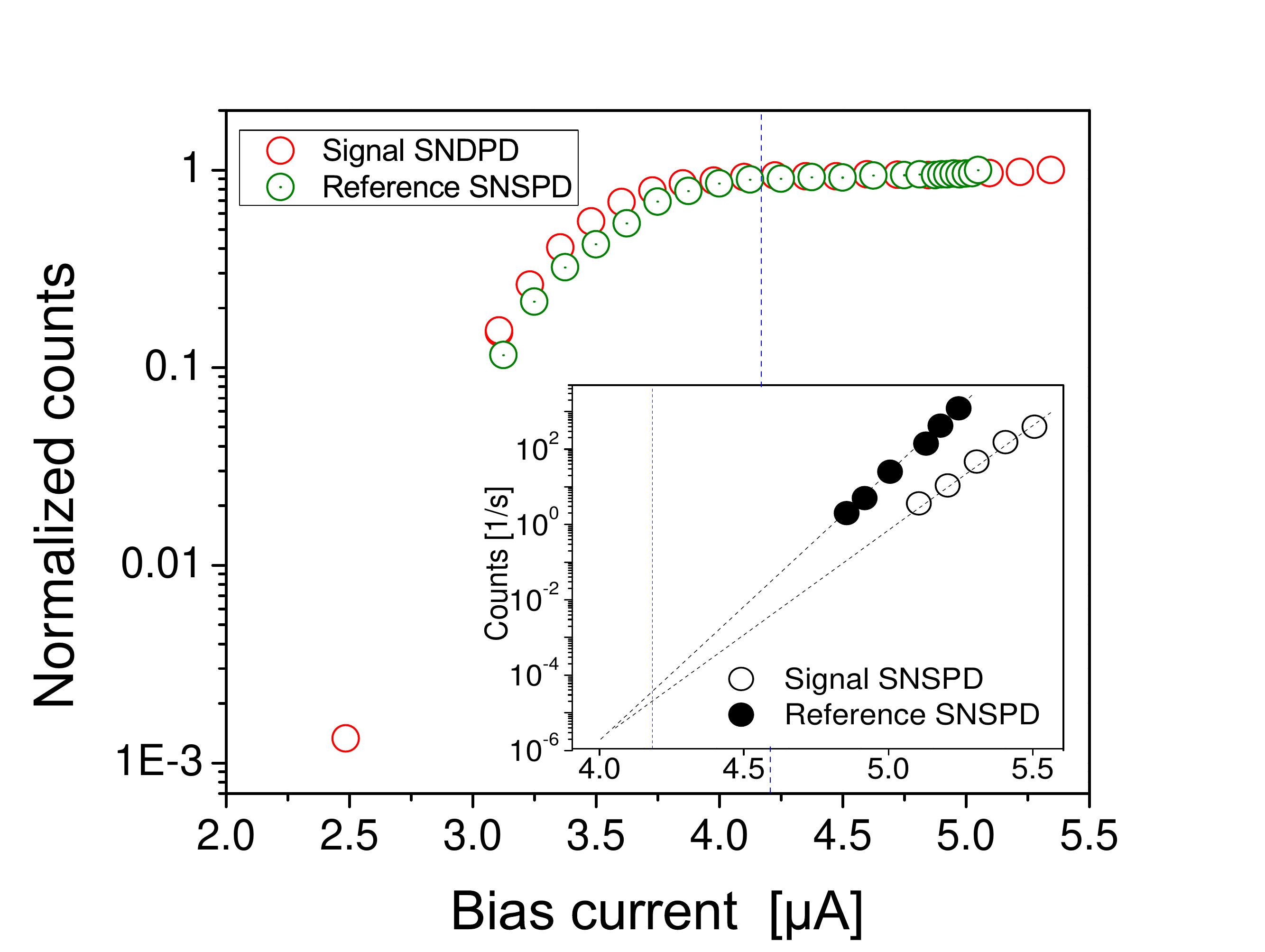}
    \caption{Normalized count rate as a function of the absolute bias current measured at 1550 nm for the primary SNSPD and reference SNSPD with identical geometry. Inset: DCR as a function of the bias current taken from both detectors.} 
    \label{fig:REF}
\end{figure}

\begin{figure}[t]
    \centering
    \includegraphics[width=0.98\columnwidth]{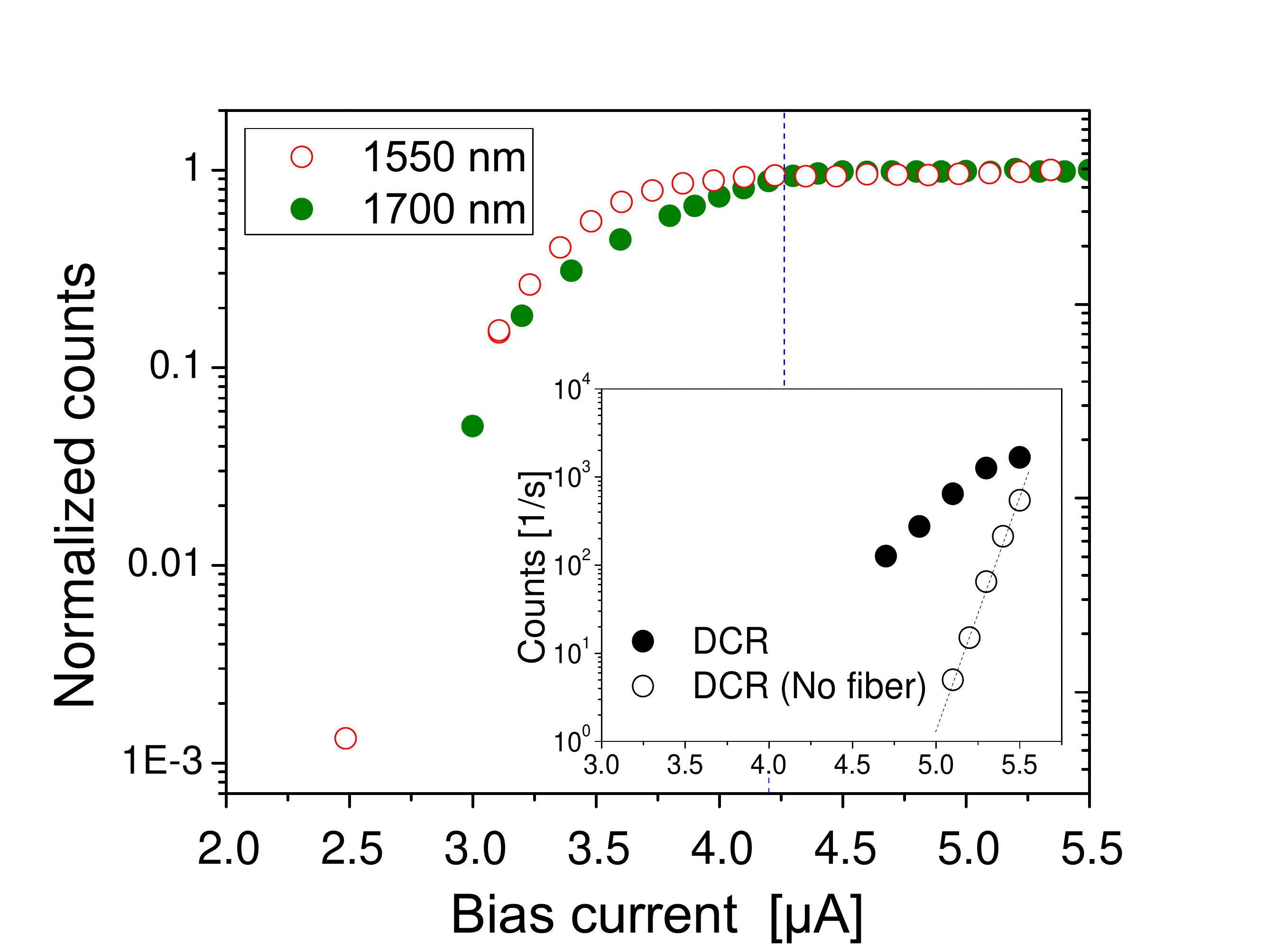}
    \caption{Normalized count rate as a function of the absolute bias current measured at 1550 nm (open red dots) and 1700 nm (green dots) wavelengths. Data was taken at 300 mK of bath temperature. The SNSPDs show pronounced saturation at both wavelengths. Inset: Comparison of DCRs with optical fiber connected (filled circles) and disconnected (open circles).} 
    \label{fig:PCR}
\end{figure}

The switching current of the detector $I_C$ was 5.5 $\mu$A
 measured at 300~mK by sweeping the current from a \SI{50}{\ohm} impedance source. Supplementary Fig.~\ref{fig:PCR} shows the dependence of the count rate on the absolute bias current for 400~by~400~$\mu$m$^2$ large-area SNSPD at 1550 nm wavelength ($\sim$0.8~eV). When the detector was
illuminated, the count rate (open red dots) rose at a bias current of 3~$\mu$A. Counts initially grew with the current and the device was nearly saturated at a bias current of 4.2~$\mu$A. At this bias current, the count rate with the laser light turned off (background count rate) was below 100 counts/s. The maximum background count rate (black dots) was measured at a point just below the switching to the normal state, at $10^3$~counts/s. The open dots represent the background count rate when the fiber was decoupled from the haloscope. Blocking of infrared photons resulted in significant reduction of the background noise. With extrapolation of the experimental data, the expected count rate at the working bias current (vertical dashed blue line) for the long-duration integration experiment is $\sim10^{-4}-10^{-5}$ Hz.  However, deviations from this extrapolation are likely when going to very low bias currents, so direct measurements of dark counts at lower bias currents is desirable in the future.

Over the total 180 hour exposure,
the experiment's reference
SNSPD registered 5 counts, while the haloscope
detector registered 4 counts,
with the latter corresponding to a count rate
of $6 \times 10^{-6} {\rm \, Hz}$. These
counts were spread fairly evenly across the integration
time, as illustrated in Supplementary Fig.~\ref{fig:long_DCR}.
As discussed in the main text, we cannot be certain as
to the cause of these events --- plausible explanations 
include cosmic ray muons,
Cherenkov photons generated in dielectric materials
such as the lens, or current fluctuations in the detector.
In particular, we do not necessarily expect the extrapolations
for the intrinsic dark count rate, discussed above
and shown in Supplementary Fig.~\ref{fig:REF},
to hold down to the bias currents used during the exposure.
Consequently, it is not surprising that the observed
dark count rates are somewhat lower than the estimates
from the previous paragraph.

Since the most plausible dark count mechanisms 
would result in almost the same dark count
rates in the reference and primary SNSPDs,
it is sensible to set limits on the signal rate
(and correspondingly the kinetic mixing parameter
$\epsilon$) assuming a common dark count rate.
Specifically, we model the reference and primary
counts as being independent and Poisson-distributed,
with means $s_d$ and $s_d + s_\epsilon$ respectively,
where $s_d$ is the expected number of dark counts
and $s_\epsilon$ is the expected number
of DM signal counts (for a given $\epsilon$).
We view $s_d$ as an unknown parameter (since we do
not have a precise estimate of the dark count rate),
and derive a 90\% limit $s_{0.9}$ on $s_\epsilon$ 
via requiring that
\begin{align}
	{\underset {s_d} {\rm max}} \, \, &\mathbb{P}\Big[
	P(n_r,n_p|s_d,s_{0.9}
	) 
	\le P(5,4|s_d,s_{0.9})  \nonumber
	\\ &\cap \, n_r \ge n_p \, | \, s_d, s_{0.9} \Big] = 0.1
\label{eq_pmax}
\end{align}
where $n_r,n_p$ are the reference and primary counts, and $P(n_r,n_p | s_d,s_\epsilon)$
is the probability of observing $(n_r,n_p)$ counts
given rates $s_d$ and $s_\epsilon$.
Since, if $s_\epsilon > s_\epsilon'$,
we have $P(n_r,n_p | s_d,s_\epsilon) < P(n_r,n_p|s_d,s_\epsilon')$
for $n_r \ge n_p$, the probability of obtaining
a limit smaller than the true value is $\le 10\%$ for 
any value of $s_d$ (where we take the limit
to be sufficiently large if we observe $n_r < n_p$).
Numerically, the LHS of Eq.~\eqref{eq_pmax} is maximized at $s_d \simeq 4.4$, 
giving $s_{0.9} \simeq 5$. We use this to derive the $90\%$ limit
on $\epsilon$ displayed in Fig. 4 of the main text.

\begin{figure}[b]
\includegraphics[width=\columnwidth]{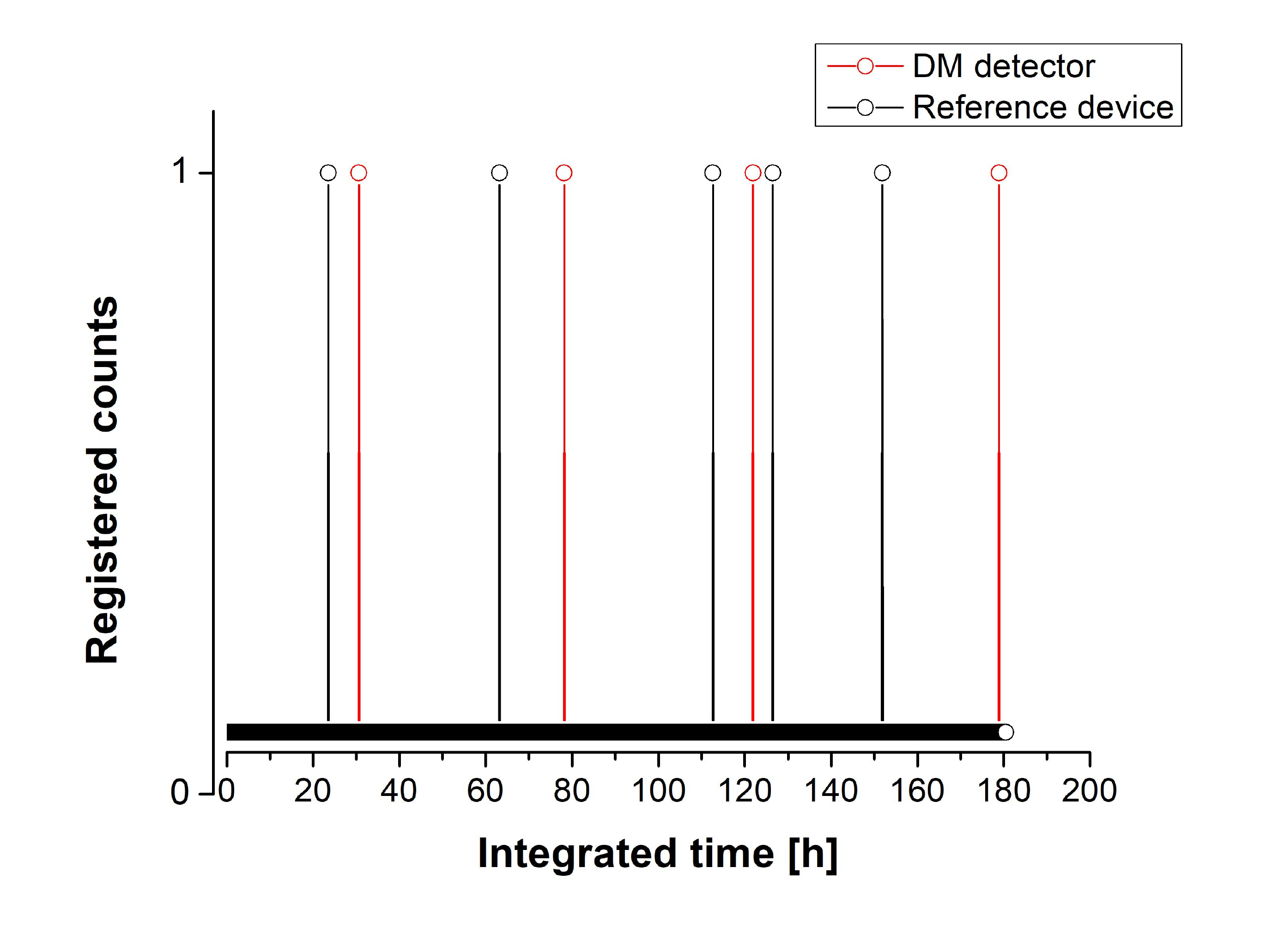}
    \caption{Experimental results of long-duration integration experiment with large-area SNSPDs mounted into a haloscope. Red points correspond to the signal obtained from a device aligned with the lens, while black points show counts taken from a reference detector placed far from focus of the target.}
      \label{fig:long_DCR}
   \end{figure}

\section{Dark photon polarization}
\label{sec_polarization}

Our limits on the coupling of dark photon dark matter are derived under the assumption that the dark photon polarization direction varies randomly over timescales longer than the DM coherence time. This is the behaviour expected from the simplest early-universe production mechanisms, such as production from inflationary fluctuations~\cite{PhysRevD.93.103520}. As discussed in~\cite{2105.04565}, it may be the case that for some other production mechanisms, the dark photon polarization direction is almost constant, over the time and length scales probed by the experiment. In such scenarios, our time-averaged power could deviate by up to $18\%$ from its value in the randomly-varying case, depending on the polarization direction. In addition, the proper statistical treatment of the noise will be slightly different, due to the time dependence of the signal photon rate.  However, these effects make only a fractionally small ($< 10\%$) difference to our limits, so we do not display separate curves.

\bibliography{darkphoton}

\begin{thebibliography}{52}%
\makeatletter
\providecommand \@ifxundefined [1]{%
 \@ifx{#1\undefined}
}%
\providecommand \@ifnum [1]{%
 \ifnum #1\expandafter \@firstoftwo
 \else \expandafter \@secondoftwo
 \fi
}%
\providecommand \@ifx [1]{%
 \ifx #1\expandafter \@firstoftwo
 \else \expandafter \@secondoftwo
 \fi
}%
\providecommand \natexlab [1]{#1}%
\providecommand \enquote  [1]{``#1''}%
\providecommand \bibnamefont  [1]{#1}%
\providecommand \bibfnamefont [1]{#1}%
\providecommand \citenamefont [1]{#1}%
\providecommand \href@noop [0]{\@secondoftwo}%
\providecommand \href [0]{\begingroup \@sanitize@url \@href}%
\providecommand \@href[1]{\@@startlink{#1}\@@href}%
\providecommand \@@href[1]{\endgroup#1\@@endlink}%
\providecommand \@sanitize@url [0]{\catcode `\\12\catcode `\$12\catcode
  `\&12\catcode `\#12\catcode `\^12\catcode `\_12\catcode `\%12\relax}%
\providecommand \@@startlink[1]{}%
\providecommand \@@endlink[0]{}%
\providecommand \url  [0]{\begingroup\@sanitize@url \@url }%
\providecommand \@url [1]{\endgroup\@href {#1}{\urlprefix }}%
\providecommand \urlprefix  [0]{URL }%
\providecommand \Eprint [0]{\href }%
\providecommand \doibase [0]{http://dx.doi.org/}%
\providecommand \selectlanguage [0]{\@gobble}%
\providecommand \bibinfo  [0]{\@secondoftwo}%
\providecommand \bibfield  [0]{\@secondoftwo}%
\providecommand \translation [1]{[#1]}%
\providecommand \BibitemOpen [0]{}%
\providecommand \bibitemStop [0]{}%
\providecommand \bibitemNoStop [0]{.\EOS\space}%
\providecommand \EOS [0]{\spacefactor3000\relax}%
\providecommand \BibitemShut  [1]{\csname bibitem#1\endcsname}%
\let\auto@bib@innerbib\@empty
\bibitem [{\citenamefont {Rubin}\ and\ \citenamefont
  {Ford}(1970)}]{Rubin:1970zza}%
  \BibitemOpen
  \bibfield  {author} {\bibinfo {author} {\bibfnamefont {V.~C.}\ \bibnamefont
  {Rubin}}\ and\ \bibinfo {author} {\bibfnamefont {W.~K.}\ \bibnamefont {Ford},
  \bibfnamefont {Jr.}},\ }\href {\doibase 10.1086/150317} {\bibfield  {journal}
  {\bibinfo  {journal} {Astrophys. J.}\ }\textbf {\bibinfo {volume} {159}},\
  \bibinfo {pages} {379} (\bibinfo {year} {1970})}\BibitemShut {NoStop}%
\bibitem [{\citenamefont {Aghanim}\ \emph {et~al.}(2020)\citenamefont {Aghanim}
  \emph {et~al.}}]{Aghanim:2018eyx}%
  \BibitemOpen
  \bibfield  {author} {\bibinfo {author} {\bibfnamefont {N.}~\bibnamefont
  {Aghanim}} \emph {et~al.} (\bibinfo {collaboration} {Planck}),\ }\href
  {\doibase 10.1051/0004-6361/201833910} {\bibfield  {journal} {\bibinfo
  {journal} {Astron. Astrophys.}\ }\textbf {\bibinfo {volume} {641}},\ \bibinfo
  {pages} {A6} (\bibinfo {year} {2020})},\ \Eprint
  {http://arxiv.org/abs/1807.06209} {arXiv:1807.06209 [astro-ph.CO]}
  \BibitemShut {NoStop}%
\bibitem [{\citenamefont {Dine}\ and\ \citenamefont
  {Fischler}(1983)}]{Dine:1982ah}%
  \BibitemOpen
  \bibfield  {author} {\bibinfo {author} {\bibfnamefont {M.}~\bibnamefont
  {Dine}}\ and\ \bibinfo {author} {\bibfnamefont {W.}~\bibnamefont
  {Fischler}},\ }\href {\doibase 10.1016/0370-2693(83)90639-1} {\bibfield
  {journal} {\bibinfo  {journal} {Phys. Lett. B}\ }\textbf {\bibinfo {volume}
  {120}},\ \bibinfo {pages} {137} (\bibinfo {year} {1983})}\BibitemShut
  {NoStop}%
\bibitem [{\citenamefont {Abbott}\ and\ \citenamefont
  {Sikivie}(1983)}]{Abbott:1982af}%
  \BibitemOpen
  \bibfield  {author} {\bibinfo {author} {\bibfnamefont {L.~F.}\ \bibnamefont
  {Abbott}}\ and\ \bibinfo {author} {\bibfnamefont {P.}~\bibnamefont
  {Sikivie}},\ }\href {\doibase 10.1016/0370-2693(83)90638-X} {\bibfield
  {journal} {\bibinfo  {journal} {Phys. Lett. B}\ }\textbf {\bibinfo {volume}
  {120}},\ \bibinfo {pages} {133} (\bibinfo {year} {1983})}\BibitemShut
  {NoStop}%
\bibitem [{\citenamefont {Preskill}\ \emph {et~al.}(1983)\citenamefont
  {Preskill}, \citenamefont {Wise},\ and\ \citenamefont
  {Wilczek}}]{PRESKILL1983127}%
  \BibitemOpen
  \bibfield  {author} {\bibinfo {author} {\bibfnamefont {J.}~\bibnamefont
  {Preskill}}, \bibinfo {author} {\bibfnamefont {M.~B.}\ \bibnamefont {Wise}},
  \ and\ \bibinfo {author} {\bibfnamefont {F.}~\bibnamefont {Wilczek}},\ }\href
  {\doibase https://doi.org/10.1016/0370-2693(83)90637-8} {\bibfield  {journal}
  {\bibinfo  {journal} {Physics Letters B}\ }\textbf {\bibinfo {volume}
  {120}},\ \bibinfo {pages} {127} (\bibinfo {year} {1983})}\BibitemShut
  {NoStop}%
\bibitem [{\citenamefont {Arias}\ \emph {et~al.}(2012)\citenamefont {Arias},
  \citenamefont {Cadamuro}, \citenamefont {Goodsell}, \citenamefont {Jaeckel},
  \citenamefont {Redondo},\ and\ \citenamefont {Ringwald}}]{Arias:2012az}%
  \BibitemOpen
  \bibfield  {author} {\bibinfo {author} {\bibfnamefont {P.}~\bibnamefont
  {Arias}}, \bibinfo {author} {\bibfnamefont {D.}~\bibnamefont {Cadamuro}},
  \bibinfo {author} {\bibfnamefont {M.}~\bibnamefont {Goodsell}}, \bibinfo
  {author} {\bibfnamefont {J.}~\bibnamefont {Jaeckel}}, \bibinfo {author}
  {\bibfnamefont {J.}~\bibnamefont {Redondo}}, \ and\ \bibinfo {author}
  {\bibfnamefont {A.}~\bibnamefont {Ringwald}},\ }\href {\doibase
  10.1088/1475-7516/2012/06/013} {\bibfield  {journal} {\bibinfo  {journal}
  {JCAP}\ }\textbf {\bibinfo {volume} {1206}},\ \bibinfo {pages} {013}
  (\bibinfo {year} {2012})},\ \Eprint {http://arxiv.org/abs/1201.5902}
  {arXiv:1201.5902 [hep-ph]} \BibitemShut {NoStop}%
\bibitem [{\citenamefont {Graham}\ \emph {et~al.}(2016)\citenamefont {Graham},
  \citenamefont {Mardon},\ and\ \citenamefont
  {Rajendran}}]{PhysRevD.93.103520}%
  \BibitemOpen
  \bibfield  {author} {\bibinfo {author} {\bibfnamefont {P.~W.}\ \bibnamefont
  {Graham}}, \bibinfo {author} {\bibfnamefont {J.}~\bibnamefont {Mardon}}, \
  and\ \bibinfo {author} {\bibfnamefont {S.}~\bibnamefont {Rajendran}},\ }\href
  {\doibase 10.1103/PhysRevD.93.103520} {\bibfield  {journal} {\bibinfo
  {journal} {Phys. Rev. D}\ }\textbf {\bibinfo {volume} {93}},\ \bibinfo
  {pages} {103520} (\bibinfo {year} {2016})}\BibitemShut {NoStop}%
\bibitem [{\citenamefont {Svrcek}\ and\ \citenamefont
  {Witten}(2006)}]{Svrcek:2006yi}%
  \BibitemOpen
  \bibfield  {author} {\bibinfo {author} {\bibfnamefont {P.}~\bibnamefont
  {Svrcek}}\ and\ \bibinfo {author} {\bibfnamefont {E.}~\bibnamefont
  {Witten}},\ }\href {\doibase 10.1088/1126-6708/2006/06/051} {\bibfield
  {journal} {\bibinfo  {journal} {JHEP}\ }\textbf {\bibinfo {volume} {06}},\
  \bibinfo {pages} {051} (\bibinfo {year} {2006})},\ \Eprint
  {http://arxiv.org/abs/hep-th/0605206} {arXiv:hep-th/0605206} \BibitemShut
  {NoStop}%
\bibitem [{\citenamefont {Arvanitaki}\ \emph {et~al.}(2010)\citenamefont
  {Arvanitaki}, \citenamefont {Dimopoulos}, \citenamefont {Dubovsky},
  \citenamefont {Kaloper},\ and\ \citenamefont
  {March-Russell}}]{Arvanitaki:2009fg}%
  \BibitemOpen
  \bibfield  {author} {\bibinfo {author} {\bibfnamefont {A.}~\bibnamefont
  {Arvanitaki}}, \bibinfo {author} {\bibfnamefont {S.}~\bibnamefont
  {Dimopoulos}}, \bibinfo {author} {\bibfnamefont {S.}~\bibnamefont
  {Dubovsky}}, \bibinfo {author} {\bibfnamefont {N.}~\bibnamefont {Kaloper}}, \
  and\ \bibinfo {author} {\bibfnamefont {J.}~\bibnamefont {March-Russell}},\
  }\href {\doibase 10.1103/PhysRevD.81.123530} {\bibfield  {journal} {\bibinfo
  {journal} {Phys. Rev. D}\ }\textbf {\bibinfo {volume} {81}},\ \bibinfo
  {pages} {123530} (\bibinfo {year} {2010})},\ \Eprint
  {http://arxiv.org/abs/0905.4720} {arXiv:0905.4720 [hep-th]} \BibitemShut
  {NoStop}%
\bibitem [{\citenamefont {Holdom}(1986)}]{Holdom:1985ag}%
  \BibitemOpen
  \bibfield  {author} {\bibinfo {author} {\bibfnamefont {B.}~\bibnamefont
  {Holdom}},\ }\href {\doibase 10.1016/0370-2693(86)91377-8} {\bibfield
  {journal} {\bibinfo  {journal} {Phys. Lett.}\ }\textbf {\bibinfo {volume}
  {B166}},\ \bibinfo {pages} {196} (\bibinfo {year} {1986})}\BibitemShut
  {NoStop}%
\bibitem [{\citenamefont {Cicoli}\ \emph {et~al.}(2011)\citenamefont {Cicoli},
  \citenamefont {Goodsell}, \citenamefont {Jaeckel},\ and\ \citenamefont
  {Ringwald}}]{Cicoli:2011yh}%
  \BibitemOpen
  \bibfield  {author} {\bibinfo {author} {\bibfnamefont {M.}~\bibnamefont
  {Cicoli}}, \bibinfo {author} {\bibfnamefont {M.}~\bibnamefont {Goodsell}},
  \bibinfo {author} {\bibfnamefont {J.}~\bibnamefont {Jaeckel}}, \ and\
  \bibinfo {author} {\bibfnamefont {A.}~\bibnamefont {Ringwald}},\ }\href
  {\doibase 10.1007/JHEP07(2011)114} {\bibfield  {journal} {\bibinfo  {journal}
  {JHEP}\ }\textbf {\bibinfo {volume} {07}},\ \bibinfo {pages} {114} (\bibinfo
  {year} {2011})},\ \Eprint {http://arxiv.org/abs/1103.3705} {arXiv:1103.3705
  [hep-th]} \BibitemShut {NoStop}%
\bibitem [{\citenamefont {Dimopoulos}\ and\ \citenamefont
  {Giudice}(1996)}]{Dimopoulos:1996kp}%
  \BibitemOpen
  \bibfield  {author} {\bibinfo {author} {\bibfnamefont {S.}~\bibnamefont
  {Dimopoulos}}\ and\ \bibinfo {author} {\bibfnamefont {G.~F.}\ \bibnamefont
  {Giudice}},\ }\href {\doibase 10.1016/0370-2693(96)00390-5} {\bibfield
  {journal} {\bibinfo  {journal} {Phys. Lett.}\ }\textbf {\bibinfo {volume}
  {B379}},\ \bibinfo {pages} {105} (\bibinfo {year} {1996})},\ \Eprint
  {http://arxiv.org/abs/hep-ph/9602350} {arXiv:hep-ph/9602350 [hep-ph]}
  \BibitemShut {NoStop}%
\bibitem [{\citenamefont {Damour}\ and\ \citenamefont
  {Polyakov}(1994)}]{Damour:1994zq}%
  \BibitemOpen
  \bibfield  {author} {\bibinfo {author} {\bibfnamefont {T.}~\bibnamefont
  {Damour}}\ and\ \bibinfo {author} {\bibfnamefont {A.~M.}\ \bibnamefont
  {Polyakov}},\ }\href {\doibase 10.1016/0550-3213(94)90143-0} {\bibfield
  {journal} {\bibinfo  {journal} {Nucl. Phys.}\ }\textbf {\bibinfo {volume}
  {B423}},\ \bibinfo {pages} {532} (\bibinfo {year} {1994})},\ \Eprint
  {http://arxiv.org/abs/hep-th/9401069} {arXiv:hep-th/9401069 [hep-th]}
  \BibitemShut {NoStop}%
\bibitem [{\citenamefont {Akrami}\ \emph {et~al.}(2020)\citenamefont {Akrami}
  \emph {et~al.}}]{Akrami:2018odb}%
  \BibitemOpen
  \bibfield  {author} {\bibinfo {author} {\bibfnamefont {Y.}~\bibnamefont
  {Akrami}} \emph {et~al.} (\bibinfo {collaboration} {Planck}),\ }\href
  {\doibase 10.1051/0004-6361/201833887} {\bibfield  {journal} {\bibinfo
  {journal} {Astron. Astrophys.}\ }\textbf {\bibinfo {volume} {641}},\ \bibinfo
  {pages} {A10} (\bibinfo {year} {2020})},\ \Eprint
  {http://arxiv.org/abs/1807.06211} {arXiv:1807.06211 [astro-ph.CO]}
  \BibitemShut {NoStop}%
\bibitem [{\citenamefont {Baryakhtar}\ \emph
  {et~al.}(2018{\natexlab{a}})\citenamefont {Baryakhtar}, \citenamefont
  {Huang},\ and\ \citenamefont {Lasenby}}]{PhysRevD.98.035006}%
  \BibitemOpen
  \bibfield  {author} {\bibinfo {author} {\bibfnamefont {M.}~\bibnamefont
  {Baryakhtar}}, \bibinfo {author} {\bibfnamefont {J.}~\bibnamefont {Huang}}, \
  and\ \bibinfo {author} {\bibfnamefont {R.}~\bibnamefont {Lasenby}},\ }\href
  {\doibase 10.1103/PhysRevD.98.035006} {\bibfield  {journal} {\bibinfo
  {journal} {Phys. Rev. D}\ }\textbf {\bibinfo {volume} {98}},\ \bibinfo
  {pages} {035006} (\bibinfo {year} {2018}{\natexlab{a}})}\BibitemShut
  {NoStop}%
\bibitem [{\citenamefont {Arvanitaki}\ \emph {et~al.}(2018)\citenamefont
  {Arvanitaki}, \citenamefont {Dimopoulos},\ and\ \citenamefont
  {Van~Tilburg}}]{PhysRevX.8.041001}%
  \BibitemOpen
  \bibfield  {author} {\bibinfo {author} {\bibfnamefont {A.}~\bibnamefont
  {Arvanitaki}}, \bibinfo {author} {\bibfnamefont {S.}~\bibnamefont
  {Dimopoulos}}, \ and\ \bibinfo {author} {\bibfnamefont {K.}~\bibnamefont
  {Van~Tilburg}},\ }\href {\doibase 10.1103/PhysRevX.8.041001} {\bibfield
  {journal} {\bibinfo  {journal} {Phys. Rev. X}\ }\textbf {\bibinfo {volume}
  {8}},\ \bibinfo {pages} {041001} (\bibinfo {year} {2018})}\BibitemShut
  {NoStop}%
\bibitem [{\citenamefont {Goodman}\ and\ \citenamefont
  {Witten}(1985)}]{Goodman:1984dc}%
  \BibitemOpen
  \bibfield  {author} {\bibinfo {author} {\bibfnamefont {M.~W.}\ \bibnamefont
  {Goodman}}\ and\ \bibinfo {author} {\bibfnamefont {E.}~\bibnamefont
  {Witten}},\ }\href {\doibase 10.1103/PhysRevD.31.3059} {\bibfield  {journal}
  {\bibinfo  {journal} {Phys. Rev. D}\ }\textbf {\bibinfo {volume} {31}},\
  \bibinfo {pages} {3059} (\bibinfo {year} {1985})}\BibitemShut {NoStop}%
\bibitem [{\citenamefont {Caldwell}\ \emph {et~al.}(2017)\citenamefont
  {Caldwell}, \citenamefont {Dvali}, \citenamefont {Majorovits}, \citenamefont
  {Millar}, \citenamefont {Raffelt}, \citenamefont {Redondo}, \citenamefont
  {Reimann}, \citenamefont {Simon},\ and\ \citenamefont
  {Steffen}}]{TheMADMAXWorkingGroup:2016hpc}%
  \BibitemOpen
  \bibfield  {author} {\bibinfo {author} {\bibfnamefont {A.}~\bibnamefont
  {Caldwell}}, \bibinfo {author} {\bibfnamefont {G.}~\bibnamefont {Dvali}},
  \bibinfo {author} {\bibfnamefont {B.}~\bibnamefont {Majorovits}}, \bibinfo
  {author} {\bibfnamefont {A.}~\bibnamefont {Millar}}, \bibinfo {author}
  {\bibfnamefont {G.}~\bibnamefont {Raffelt}}, \bibinfo {author} {\bibfnamefont
  {J.}~\bibnamefont {Redondo}}, \bibinfo {author} {\bibfnamefont
  {O.}~\bibnamefont {Reimann}}, \bibinfo {author} {\bibfnamefont
  {F.}~\bibnamefont {Simon}}, \ and\ \bibinfo {author} {\bibfnamefont
  {F.}~\bibnamefont {Steffen}} (\bibinfo {collaboration} {MADMAX Working
  Group}),\ }\href {\doibase 10.1103/PhysRevLett.118.091801} {\bibfield
  {journal} {\bibinfo  {journal} {Phys. Rev. Lett.}\ }\textbf {\bibinfo
  {volume} {118}},\ \bibinfo {pages} {091801} (\bibinfo {year} {2017})},\
  \Eprint {http://arxiv.org/abs/1611.05865} {arXiv:1611.05865
  [physics.ins-det]} \BibitemShut {NoStop}%
\bibitem [{\citenamefont {Carosi}\ \emph {et~al.}(2020)\citenamefont {Carosi},
  \citenamefont {Cervantes}, \citenamefont {Kimes}, \citenamefont {Mohapatra},
  \citenamefont {Ottens},\ and\ \citenamefont {Rybka}}]{orpheus}%
  \BibitemOpen
  \bibfield  {author} {\bibinfo {author} {\bibfnamefont {G.}~\bibnamefont
  {Carosi}}, \bibinfo {author} {\bibfnamefont {R.}~\bibnamefont {Cervantes}},
  \bibinfo {author} {\bibfnamefont {S.}~\bibnamefont {Kimes}}, \bibinfo
  {author} {\bibfnamefont {P.}~\bibnamefont {Mohapatra}}, \bibinfo {author}
  {\bibfnamefont {R.}~\bibnamefont {Ottens}}, \ and\ \bibinfo {author}
  {\bibfnamefont {G.}~\bibnamefont {Rybka}},\ }in\ \href@noop {} {\emph
  {\bibinfo {booktitle} {Microwave Cavities and Detectors for Axion
  Research}}},\ \bibinfo {editor} {edited by\ \bibinfo {editor} {\bibfnamefont
  {G.}~\bibnamefont {Carosi}}\ and\ \bibinfo {editor} {\bibfnamefont
  {G.}~\bibnamefont {Rybka}}}\ (\bibinfo  {publisher} {Springer International
  Publishing},\ \bibinfo {address} {Cham},\ \bibinfo {year} {2020})\ pp.\
  \bibinfo {pages} {169--175}\BibitemShut {NoStop}%
\bibitem [{\citenamefont {Phillips}(2017)}]{etiger1}%
  \BibitemOpen
  \bibfield  {author} {\bibinfo {author} {\bibfnamefont {B.}~\bibnamefont
  {Phillips}},\ }in\ \href@noop {} {\emph {\bibinfo {booktitle} {{2nd Workshop
  on Microwave Cavities and Detectors for Axion Research, Lawrence Livermore
  National Laboratory}}}}\ (\bibinfo {year} {2017})\BibitemShut {NoStop}%
\bibitem [{\citenamefont {Sloan}(2015)}]{etiger2}%
  \BibitemOpen
  \bibfield  {author} {\bibinfo {author} {\bibfnamefont {J.}~\bibnamefont
  {Sloan}},\ }in\ \href@noop {} {\emph {\bibinfo {booktitle} {{1nd Workshop on
  Microwave Cavity Design for Axion detection, Livermore Valley Open
  Campus}}}}\ (\bibinfo {year} {2015})\BibitemShut {NoStop}%
\bibitem [{\citenamefont {Baryakhtar}\ \emph
  {et~al.}(2018{\natexlab{b}})\citenamefont {Baryakhtar}, \citenamefont
  {Huang},\ and\ \citenamefont {Lasenby}}]{Baryakhtar:2018doz}%
  \BibitemOpen
  \bibfield  {author} {\bibinfo {author} {\bibfnamefont {M.}~\bibnamefont
  {Baryakhtar}}, \bibinfo {author} {\bibfnamefont {J.}~\bibnamefont {Huang}}, \
  and\ \bibinfo {author} {\bibfnamefont {R.}~\bibnamefont {Lasenby}},\ }\href
  {\doibase 10.1103/PhysRevD.98.035006} {\bibfield  {journal} {\bibinfo
  {journal} {Phys. Rev.}\ }\textbf {\bibinfo {volume} {D98}},\ \bibinfo {pages}
  {035006} (\bibinfo {year} {2018}{\natexlab{b}})},\ \Eprint
  {http://arxiv.org/abs/1803.11455} {arXiv:1803.11455 [hep-ph]} \BibitemShut
  {NoStop}%
\bibitem [{\citenamefont {Hochberg}\ \emph {et~al.}(2019)\citenamefont
  {Hochberg}, \citenamefont {Charaev}, \citenamefont {Nam}, \citenamefont
  {Verma}, \citenamefont {Colangelo},\ and\ \citenamefont
  {Berggren}}]{PhysRevLett.123.151802}%
  \BibitemOpen
  \bibfield  {author} {\bibinfo {author} {\bibfnamefont {Y.}~\bibnamefont
  {Hochberg}}, \bibinfo {author} {\bibfnamefont {I.}~\bibnamefont {Charaev}},
  \bibinfo {author} {\bibfnamefont {S.-W.}\ \bibnamefont {Nam}}, \bibinfo
  {author} {\bibfnamefont {V.}~\bibnamefont {Verma}}, \bibinfo {author}
  {\bibfnamefont {M.}~\bibnamefont {Colangelo}}, \ and\ \bibinfo {author}
  {\bibfnamefont {K.~K.}\ \bibnamefont {Berggren}},\ }\href {\doibase
  10.1103/PhysRevLett.123.151802} {\bibfield  {journal} {\bibinfo  {journal}
  {Phys. Rev. Lett.}\ }\textbf {\bibinfo {volume} {123}},\ \bibinfo {pages}
  {151802} (\bibinfo {year} {2019})}\BibitemShut {NoStop}%
\bibitem [{\citenamefont {Reddy}\ \emph {et~al.}(2020)\citenamefont {Reddy},
  \citenamefont {Nerem}, \citenamefont {Nam}, \citenamefont {Mirin},\ and\
  \citenamefont {Verma}}]{Reddy20}%
  \BibitemOpen
  \bibfield  {author} {\bibinfo {author} {\bibfnamefont {D.~V.}\ \bibnamefont
  {Reddy}}, \bibinfo {author} {\bibfnamefont {R.~R.}\ \bibnamefont {Nerem}},
  \bibinfo {author} {\bibfnamefont {S.~W.}\ \bibnamefont {Nam}}, \bibinfo
  {author} {\bibfnamefont {R.~P.}\ \bibnamefont {Mirin}}, \ and\ \bibinfo
  {author} {\bibfnamefont {V.~B.}\ \bibnamefont {Verma}},\ }\href {\doibase
  10.1364/OPTICA.400751} {\bibfield  {journal} {\bibinfo  {journal} {Optica}\
  }\textbf {\bibinfo {volume} {7}},\ \bibinfo {pages} {1649} (\bibinfo {year}
  {2020})}\BibitemShut {NoStop}%
\bibitem [{\citenamefont {Verma}\ \emph {et~al.}(2020)\citenamefont {Verma}
  \emph {et~al.}}]{verma2020singlephoton}%
  \BibitemOpen
  \bibfield  {author} {\bibinfo {author} {\bibfnamefont {V.~B.}\ \bibnamefont
  {Verma}} \emph {et~al.},\ }\href@noop {} {\enquote {\bibinfo {title}
  {Single-photon detection in the mid-infrared up to 10 micron wavelength using
  tungsten silicide superconducting nanowire detectors},}\ } (\bibinfo {year}
  {2020}),\ \Eprint {http://arxiv.org/abs/2012.09979} {arXiv:2012.09979
  [physics.ins-det]} \BibitemShut {NoStop}%
\bibitem [{\citenamefont {Wollman}\ \emph {et~al.}(2017)\citenamefont {Wollman}
  \emph {et~al.}}]{Wollman17}%
  \BibitemOpen
  \bibfield  {author} {\bibinfo {author} {\bibfnamefont {E.~E.}\ \bibnamefont
  {Wollman}} \emph {et~al.},\ }\href {\doibase 10.1364/OE.25.026792} {\bibfield
   {journal} {\bibinfo  {journal} {Opt. Express}\ }\textbf {\bibinfo {volume}
  {25}},\ \bibinfo {pages} {26792} (\bibinfo {year} {2017})}\BibitemShut
  {NoStop}%
\bibitem [{\citenamefont {Du}\ \emph {et~al.}(2020)\citenamefont {Du},
  \citenamefont {Egana-Ugrinovic}, \citenamefont {Essig},\ and\ \citenamefont
  {Sholapurkar}}]{du2020sources}%
  \BibitemOpen
  \bibfield  {author} {\bibinfo {author} {\bibfnamefont {P.}~\bibnamefont
  {Du}}, \bibinfo {author} {\bibfnamefont {D.}~\bibnamefont {Egana-Ugrinovic}},
  \bibinfo {author} {\bibfnamefont {R.}~\bibnamefont {Essig}}, \ and\ \bibinfo
  {author} {\bibfnamefont {M.}~\bibnamefont {Sholapurkar}},\ }\href@noop {}
  {\enquote {\bibinfo {title} {Sources of low-energy events in low-threshold
  dark matter detectors},}\ } (\bibinfo {year} {2020}),\ \Eprint
  {http://arxiv.org/abs/2011.13939} {arXiv:2011.13939 [hep-ph]} \BibitemShut
  {NoStop}%
\bibitem [{\citenamefont {Andrianavalomahefa}\ \emph
  {et~al.}(2020)\citenamefont {Andrianavalomahefa} \emph
  {et~al.}}]{PhysRevD.102.042001}%
  \BibitemOpen
  \bibfield  {author} {\bibinfo {author} {\bibfnamefont {A.}~\bibnamefont
  {Andrianavalomahefa}} \emph {et~al.} (\bibinfo {collaboration} {The FUNK
  Experiment}),\ }\href {\doibase 10.1103/PhysRevD.102.042001} {\bibfield
  {journal} {\bibinfo  {journal} {Phys. Rev. D}\ }\textbf {\bibinfo {volume}
  {102}},\ \bibinfo {pages} {042001} (\bibinfo {year} {2020})}\BibitemShut
  {NoStop}%
\bibitem [{\citenamefont {Barak}\ \emph {et~al.}(2020)\citenamefont {Barak}
  \emph {et~al.}}]{PhysRevLett.125.171802}%
  \BibitemOpen
  \bibfield  {author} {\bibinfo {author} {\bibfnamefont {L.}~\bibnamefont
  {Barak}} \emph {et~al.} (\bibinfo {collaboration} {SENSEI Collaboration}),\
  }\href {\doibase 10.1103/PhysRevLett.125.171802} {\bibfield  {journal}
  {\bibinfo  {journal} {Phys. Rev. Lett.}\ }\textbf {\bibinfo {volume} {125}},\
  \bibinfo {pages} {171802} (\bibinfo {year} {2020})}\BibitemShut {NoStop}%
\bibitem [{\citenamefont {Angle}\ \emph {et~al.}(2011)\citenamefont {Angle}
  \emph {et~al.}}]{PhysRevLett.107.051301}%
  \BibitemOpen
  \bibfield  {author} {\bibinfo {author} {\bibfnamefont {J.}~\bibnamefont
  {Angle}} \emph {et~al.} (\bibinfo {collaboration} {XENON10 Collaboration}),\
  }\href {\doibase 10.1103/PhysRevLett.107.051301} {\bibfield  {journal}
  {\bibinfo  {journal} {Phys. Rev. Lett.}\ }\textbf {\bibinfo {volume} {107}},\
  \bibinfo {pages} {051301} (\bibinfo {year} {2011})}\BibitemShut {NoStop}%
\bibitem [{\citenamefont {An}\ \emph {et~al.}(2020)\citenamefont {An},
  \citenamefont {Pospelov}, \citenamefont {Pradler},\ and\ \citenamefont
  {Ritz}}]{PhysRevD.102.115022}%
  \BibitemOpen
  \bibfield  {author} {\bibinfo {author} {\bibfnamefont {H.}~\bibnamefont
  {An}}, \bibinfo {author} {\bibfnamefont {M.}~\bibnamefont {Pospelov}},
  \bibinfo {author} {\bibfnamefont {J.}~\bibnamefont {Pradler}}, \ and\
  \bibinfo {author} {\bibfnamefont {A.}~\bibnamefont {Ritz}},\ }\href {\doibase
  10.1103/PhysRevD.102.115022} {\bibfield  {journal} {\bibinfo  {journal}
  {Phys. Rev. D}\ }\textbf {\bibinfo {volume} {102}},\ \bibinfo {pages}
  {115022} (\bibinfo {year} {2020})}\BibitemShut {NoStop}%
\bibitem [{\citenamefont {McCabe}(2010)}]{10.1103/PhysRevD.82.023530}%
  \BibitemOpen
  \bibfield  {author} {\bibinfo {author} {\bibfnamefont {C.}~\bibnamefont
  {McCabe}},\ }\href {\doibase 10.1103/physrevd.82.023530} {\bibfield
  {journal} {\bibinfo  {journal} {Physical Review D}\ }\textbf {\bibinfo
  {volume} {82}} (\bibinfo {year} {2010}),\
  10.1103/physrevd.82.023530}\BibitemShut {NoStop}%
\bibitem [{\citenamefont {Millar}\ \emph {et~al.}(2017)\citenamefont {Millar},
  \citenamefont {Raffelt}, \citenamefont {Redondo},\ and\ \citenamefont
  {Steffen}}]{1612.07057}%
  \BibitemOpen
  \bibfield  {author} {\bibinfo {author} {\bibfnamefont {A.~J.}\ \bibnamefont
  {Millar}}, \bibinfo {author} {\bibfnamefont {G.~G.}\ \bibnamefont {Raffelt}},
  \bibinfo {author} {\bibfnamefont {J.}~\bibnamefont {Redondo}}, \ and\
  \bibinfo {author} {\bibfnamefont {F.~D.}\ \bibnamefont {Steffen}},\ }\href
  {\doibase 10.1088/1475-7516/2017/01/061} {\bibfield  {journal} {\bibinfo
  {journal} {JCAP}\ }\textbf {\bibinfo {volume} {01}},\ \bibinfo {pages} {061}
  (\bibinfo {year} {2017})},\ \Eprint {http://arxiv.org/abs/1612.07057}
  {arXiv:1612.07057 [hep-ph]} \BibitemShut {NoStop}%
\bibitem [{\citenamefont {Horns}\ \emph {et~al.}(2013)\citenamefont {Horns},
  \citenamefont {Jaeckel}, \citenamefont {Lindner}, \citenamefont {Lobanov},
  \citenamefont {Redondo},\ and\ \citenamefont {Ringwald}}]{Horns:2012jf}%
  \BibitemOpen
  \bibfield  {author} {\bibinfo {author} {\bibfnamefont {D.}~\bibnamefont
  {Horns}}, \bibinfo {author} {\bibfnamefont {J.}~\bibnamefont {Jaeckel}},
  \bibinfo {author} {\bibfnamefont {A.}~\bibnamefont {Lindner}}, \bibinfo
  {author} {\bibfnamefont {A.}~\bibnamefont {Lobanov}}, \bibinfo {author}
  {\bibfnamefont {J.}~\bibnamefont {Redondo}}, \ and\ \bibinfo {author}
  {\bibfnamefont {A.}~\bibnamefont {Ringwald}},\ }\href {\doibase
  10.1088/1475-7516/2013/04/016} {\bibfield  {journal} {\bibinfo  {journal}
  {JCAP}\ }\textbf {\bibinfo {volume} {1304}},\ \bibinfo {pages} {016}
  (\bibinfo {year} {2013})},\ \Eprint {http://arxiv.org/abs/1212.2970}
  {arXiv:1212.2970 [hep-ph]} \BibitemShut {NoStop}%
\bibitem [{\citenamefont {Suzuki}\ \emph {et~al.}(2015)\citenamefont {Suzuki},
  \citenamefont {Horie}, \citenamefont {Inoue},\ and\ \citenamefont
  {Minowa}}]{Suzuki:2015sza}%
  \BibitemOpen
  \bibfield  {author} {\bibinfo {author} {\bibfnamefont {J.}~\bibnamefont
  {Suzuki}}, \bibinfo {author} {\bibfnamefont {T.}~\bibnamefont {Horie}},
  \bibinfo {author} {\bibfnamefont {Y.}~\bibnamefont {Inoue}}, \ and\ \bibinfo
  {author} {\bibfnamefont {M.}~\bibnamefont {Minowa}},\ }\href {\doibase
  10.1088/1475-7516/2015/09/042, 10.1088/1475-7516/2015/9/042} {\bibfield
  {journal} {\bibinfo  {journal} {JCAP}\ }\textbf {\bibinfo {volume} {1509}},\
  \bibinfo {pages} {042} (\bibinfo {year} {2015})},\ \Eprint
  {http://arxiv.org/abs/1504.00118} {arXiv:1504.00118 [hep-ex]} \BibitemShut
  {NoStop}%
\bibitem [{\citenamefont {Marsili}\ \emph {et~al.}(2012)\citenamefont
  {Marsili}, \citenamefont {Bellei}, \citenamefont {Najafi}, \citenamefont
  {Dane}, \citenamefont {Dauler}, \citenamefont {Molnar},\ and\ \citenamefont
  {Berggren}}]{marsili2012efficient}%
  \BibitemOpen
  \bibfield  {author} {\bibinfo {author} {\bibfnamefont {F.}~\bibnamefont
  {Marsili}}, \bibinfo {author} {\bibfnamefont {F.}~\bibnamefont {Bellei}},
  \bibinfo {author} {\bibfnamefont {F.}~\bibnamefont {Najafi}}, \bibinfo
  {author} {\bibfnamefont {A.~E.}\ \bibnamefont {Dane}}, \bibinfo {author}
  {\bibfnamefont {E.~A.}\ \bibnamefont {Dauler}}, \bibinfo {author}
  {\bibfnamefont {R.~J.}\ \bibnamefont {Molnar}}, \ and\ \bibinfo {author}
  {\bibfnamefont {K.~K.}\ \bibnamefont {Berggren}},\ }\href@noop {} {\bibfield
  {journal} {\bibinfo  {journal} {Nano letters}\ }\textbf {\bibinfo {volume}
  {12}},\ \bibinfo {pages} {4799} (\bibinfo {year} {2012})}\BibitemShut
  {NoStop}%
\bibitem [{\citenamefont {Weinberg}(1978)}]{axion1}%
  \BibitemOpen
  \bibfield  {author} {\bibinfo {author} {\bibfnamefont {S.}~\bibnamefont
  {Weinberg}},\ }\href {\doibase 10.1103/PhysRevLett.40.223} {\bibfield
  {journal} {\bibinfo  {journal} {Phys.Rev.Lett.}\ }\textbf {\bibinfo {volume}
  {40}},\ \bibinfo {pages} {223} (\bibinfo {year} {1978})}\BibitemShut
  {NoStop}%
\bibitem [{\citenamefont {Wilczek}(1978)}]{axion2}%
  \BibitemOpen
  \bibfield  {author} {\bibinfo {author} {\bibfnamefont {F.}~\bibnamefont
  {Wilczek}},\ }\href {\doibase 10.1103/PhysRevLett.40.279} {\bibfield
  {journal} {\bibinfo  {journal} {Phys.Rev.Lett.}\ }\textbf {\bibinfo {volume}
  {40}},\ \bibinfo {pages} {279} (\bibinfo {year} {1978})}\BibitemShut
  {NoStop}%
\bibitem [{\citenamefont {Peccei}\ and\ \citenamefont {Quinn}(1977)}]{axion3}%
  \BibitemOpen
  \bibfield  {author} {\bibinfo {author} {\bibfnamefont {R.}~\bibnamefont
  {Peccei}}\ and\ \bibinfo {author} {\bibfnamefont {H.~R.}\ \bibnamefont
  {Quinn}},\ }\href {\doibase 10.1103/PhysRevLett.38.1440} {\bibfield
  {journal} {\bibinfo  {journal} {Phys.Rev.Lett.}\ }\textbf {\bibinfo {volume}
  {38}},\ \bibinfo {pages} {1440} (\bibinfo {year} {1977})}\BibitemShut
  {NoStop}%
\bibitem [{\citenamefont {Graham}\ \emph {et~al.}(2015)\citenamefont {Graham},
  \citenamefont {Irastorza}, \citenamefont {Lamoreaux}, \citenamefont
  {Lindner},\ and\ \citenamefont {van
  Bibber}}]{10.1146/annurev-nucl-102014-022120}%
  \BibitemOpen
  \bibfield  {author} {\bibinfo {author} {\bibfnamefont {P.~W.}\ \bibnamefont
  {Graham}}, \bibinfo {author} {\bibfnamefont {I.~G.}\ \bibnamefont
  {Irastorza}}, \bibinfo {author} {\bibfnamefont {S.~K.}\ \bibnamefont
  {Lamoreaux}}, \bibinfo {author} {\bibfnamefont {A.}~\bibnamefont {Lindner}},
  \ and\ \bibinfo {author} {\bibfnamefont {K.~A.}\ \bibnamefont {van Bibber}},\
  }\href {\doibase 10.1146/annurev-nucl-102014-022120} {\bibfield  {journal}
  {\bibinfo  {journal} {Annual Review of Nuclear and Particle Science}\
  }\textbf {\bibinfo {volume} {65}},\ \bibinfo {pages} {485} (\bibinfo {year}
  {2015})}\BibitemShut {NoStop}%
\bibitem [{\citenamefont {Polakovic}\ \emph {et~al.}(2020)\citenamefont
  {Polakovic}, \citenamefont {Armstrong}, \citenamefont {Yefremenko},
  \citenamefont {Pearson}, \citenamefont {Hafidi}, \citenamefont {Karapetrov},
  \citenamefont {Meziani},\ and\ \citenamefont
  {Novosad}}]{POLAKOVIC2020163543}%
  \BibitemOpen
  \bibfield  {author} {\bibinfo {author} {\bibfnamefont {T.}~\bibnamefont
  {Polakovic}}, \bibinfo {author} {\bibfnamefont {W.}~\bibnamefont
  {Armstrong}}, \bibinfo {author} {\bibfnamefont {V.}~\bibnamefont
  {Yefremenko}}, \bibinfo {author} {\bibfnamefont {J.}~\bibnamefont {Pearson}},
  \bibinfo {author} {\bibfnamefont {K.}~\bibnamefont {Hafidi}}, \bibinfo
  {author} {\bibfnamefont {G.}~\bibnamefont {Karapetrov}}, \bibinfo {author}
  {\bibfnamefont {Z.-E.}\ \bibnamefont {Meziani}}, \ and\ \bibinfo {author}
  {\bibfnamefont {V.}~\bibnamefont {Novosad}},\ }\href {\doibase
  https://doi.org/10.1016/j.nima.2020.163543} {\bibfield  {journal} {\bibinfo
  {journal} {Nuclear Instruments and Methods in Physics Research Section A:
  Accelerators, Spectrometers, Detectors and Associated Equipment}\ }\textbf
  {\bibinfo {volume} {959}},\ \bibinfo {pages} {163543} (\bibinfo {year}
  {2020})}\BibitemShut {NoStop}%
\bibitem [{\citenamefont {Lawrie}\ \emph {et~al.}(2021)\citenamefont {Lawrie},
  \citenamefont {Marvinney}, \citenamefont {Pai}, \citenamefont {Feldman},
  \citenamefont {Zhang}, \citenamefont {Miller}, \citenamefont {Hua},
  \citenamefont {Dumitrescu},\ and\ \citenamefont
  {Halász}}]{lawrie2021multifunctional}%
  \BibitemOpen
  \bibfield  {author} {\bibinfo {author} {\bibfnamefont {B.~J.}\ \bibnamefont
  {Lawrie}}, \bibinfo {author} {\bibfnamefont {C.~E.}\ \bibnamefont
  {Marvinney}}, \bibinfo {author} {\bibfnamefont {Y.-Y.}\ \bibnamefont {Pai}},
  \bibinfo {author} {\bibfnamefont {M.~A.}\ \bibnamefont {Feldman}}, \bibinfo
  {author} {\bibfnamefont {J.}~\bibnamefont {Zhang}}, \bibinfo {author}
  {\bibfnamefont {A.~J.}\ \bibnamefont {Miller}}, \bibinfo {author}
  {\bibfnamefont {C.}~\bibnamefont {Hua}}, \bibinfo {author} {\bibfnamefont
  {E.}~\bibnamefont {Dumitrescu}}, \ and\ \bibinfo {author} {\bibfnamefont
  {G.~B.}\ \bibnamefont {Halász}},\ }\href@noop {} {\enquote {\bibinfo {title}
  {Multifunctional superconducting nanowire quantum sensors},}\ } (\bibinfo
  {year} {2021}),\ \Eprint {http://arxiv.org/abs/2103.09896} {arXiv:2103.09896
  [quant-ph]} \BibitemShut {NoStop}%
\bibitem [{\citenamefont {Moharam}\ and\ \citenamefont
  {Gaylord}(1981)}]{Moharam:81}%
  \BibitemOpen
  \bibfield  {author} {\bibinfo {author} {\bibfnamefont {M.~G.}\ \bibnamefont
  {Moharam}}\ and\ \bibinfo {author} {\bibfnamefont {T.~K.}\ \bibnamefont
  {Gaylord}},\ }\href {\doibase 10.1364/JOSA.71.000811} {\bibfield  {journal}
  {\bibinfo  {journal} {J. Opt. Soc. Am.}\ }\textbf {\bibinfo {volume} {71}},\
  \bibinfo {pages} {811} (\bibinfo {year} {1981})}\BibitemShut {NoStop}%
\bibitem [{\citenamefont {Pedrotti}\ \emph {et~al.}(2017)\citenamefont
  {Pedrotti}, \citenamefont {Pedrotti},\ and\ \citenamefont
  {Pedrotti}}]{pedrotti2017introduction}%
  \BibitemOpen
  \bibfield  {author} {\bibinfo {author} {\bibfnamefont {F.~L.}\ \bibnamefont
  {Pedrotti}}, \bibinfo {author} {\bibfnamefont {L.~M.}\ \bibnamefont
  {Pedrotti}}, \ and\ \bibinfo {author} {\bibfnamefont {L.~S.}\ \bibnamefont
  {Pedrotti}},\ }\href@noop {} {\emph {\bibinfo {title} {Introduction to
  optics}}}\ (\bibinfo  {publisher} {Cambridge University Press},\ \bibinfo
  {year} {2017})\BibitemShut {NoStop}%
\bibitem [{\citenamefont {Lyon}\ \emph {et~al.}(1977)\citenamefont {Lyon},
  \citenamefont {Salinger}, \citenamefont {Swenson},\ and\ \citenamefont
  {White}}]{Lyon1977ThermalExpSi}%
  \BibitemOpen
  \bibfield  {author} {\bibinfo {author} {\bibfnamefont {K.~G.}\ \bibnamefont
  {Lyon}}, \bibinfo {author} {\bibfnamefont {G.~L.}\ \bibnamefont {Salinger}},
  \bibinfo {author} {\bibfnamefont {C.~A.}\ \bibnamefont {Swenson}}, \ and\
  \bibinfo {author} {\bibfnamefont {G.~K.}\ \bibnamefont {White}},\ }\href
  {\doibase 10.1063/1.323747} {\bibfield  {journal} {\bibinfo  {journal}
  {Journal of Applied Physics}\ }\textbf {\bibinfo {volume} {48}},\ \bibinfo
  {pages} {865} (\bibinfo {year} {1977})},\ \Eprint
  {http://arxiv.org/abs/https://doi.org/10.1063/1.323747}
  {https://doi.org/10.1063/1.323747} \BibitemShut {NoStop}%
\bibitem [{\citenamefont {Hahn}\ and\ \citenamefont
  {Kirby}(1972)}]{Hahn1972ThermalExpSio2}%
  \BibitemOpen
  \bibfield  {author} {\bibinfo {author} {\bibfnamefont {T.~A.}\ \bibnamefont
  {Hahn}}\ and\ \bibinfo {author} {\bibfnamefont {R.~K.}\ \bibnamefont
  {Kirby}},\ }\href {\doibase 10.1063/1.2948551} {\bibfield  {journal}
  {\bibinfo  {journal} {AIP Conference Proceedings}\ }\textbf {\bibinfo
  {volume} {3}},\ \bibinfo {pages} {13} (\bibinfo {year} {1972})},\ \Eprint
  {http://arxiv.org/abs/https://aip.scitation.org/doi/pdf/10.1063/1.2948551}
  {https://aip.scitation.org/doi/pdf/10.1063/1.2948551} \BibitemShut {NoStop}%
\bibitem [{vla()}]{vlassak}%
  \BibitemOpen
  \href@noop {} {\enquote {\bibinfo {title} {Thin film mechanics},}\ }\bibinfo
  {howpublished}
  {\url{https://www.mrsec.harvard.edu/education/ap298r2004/Vlassak\%20AP298presentation.pdf}}\BibitemShut
  {NoStop}%
\bibitem [{\citenamefont {Stoney}\ and\ \citenamefont
  {Parsons}(1909)}]{Stoney1909}%
  \BibitemOpen
  \bibfield  {author} {\bibinfo {author} {\bibfnamefont {G.~G.}\ \bibnamefont
  {Stoney}}\ and\ \bibinfo {author} {\bibfnamefont {C.~A.}\ \bibnamefont
  {Parsons}},\ }\href {\doibase 10.1098/rspa.1909.0021} {\bibfield  {journal}
  {\bibinfo  {journal} {Proceedings of the Royal Society of London. Series A,
  Containing Papers of a Mathematical and Physical Character}\ }\textbf
  {\bibinfo {volume} {82}},\ \bibinfo {pages} {172} (\bibinfo {year}
  {1909})}\BibitemShut {NoStop}%
\bibitem [{wik()}]{wikipedia_2021}%
  \BibitemOpen
  \href@noop {} {\enquote {\bibinfo {title} {Lambert's cosine law},}\ }\bibinfo
  {howpublished}
  {\url{https://en.wikipedia.org/wiki/Lambert's_cosine_law}}\BibitemShut
  {NoStop}%
\bibitem [{\citenamefont {Komma}\ \emph {et~al.}(2012)\citenamefont {Komma},
  \citenamefont {Schwarz}, \citenamefont {Hofmann}, \citenamefont {Heinert},\
  and\ \citenamefont {Nawrodt}}]{Komma2012ThermoOptic}%
  \BibitemOpen
  \bibfield  {author} {\bibinfo {author} {\bibfnamefont {J.}~\bibnamefont
  {Komma}}, \bibinfo {author} {\bibfnamefont {C.}~\bibnamefont {Schwarz}},
  \bibinfo {author} {\bibfnamefont {G.}~\bibnamefont {Hofmann}}, \bibinfo
  {author} {\bibfnamefont {D.}~\bibnamefont {Heinert}}, \ and\ \bibinfo
  {author} {\bibfnamefont {R.}~\bibnamefont {Nawrodt}},\ }\href {\doibase
  10.1063/1.4738989} {\bibfield  {journal} {\bibinfo  {journal} {Applied
  Physics Letters}\ }\textbf {\bibinfo {volume} {101}},\ \bibinfo {pages}
  {041905} (\bibinfo {year} {2012})},\ \Eprint
  {http://arxiv.org/abs/https://doi.org/10.1063/1.4738989}
  {https://doi.org/10.1063/1.4738989} \BibitemShut {NoStop}%
\bibitem [{\citenamefont {Caputo}\ \emph {et~al.}(2021)\citenamefont {Caputo},
  \citenamefont {Millar}, \citenamefont {O'Hare},\ and\ \citenamefont
  {Vitagliano}}]{2105.04565}%
  \BibitemOpen
  \bibfield  {author} {\bibinfo {author} {\bibfnamefont {A.}~\bibnamefont
  {Caputo}}, \bibinfo {author} {\bibfnamefont {A.~J.}\ \bibnamefont {Millar}},
  \bibinfo {author} {\bibfnamefont {C.~A.~J.}\ \bibnamefont {O'Hare}}, \ and\
  \bibinfo {author} {\bibfnamefont {E.}~\bibnamefont {Vitagliano}},\
  }\href@noop {} {\  (\bibinfo {year} {2021})},\ \Eprint
  {http://arxiv.org/abs/2105.04565} {arXiv:2105.04565 [hep-ph]} \BibitemShut
  {NoStop}%
\bibitem [{Sup()}]{Supp}%
  \BibitemOpen
  \href@noop {} {}\bibinfo {note} {See Supplemental Material for additional
  experimental and theoretical details, which includes Refs.
  [43-51].}\BibitemShut {Stop}%
\end{thebibliography}%
\end{document}